\documentclass[a4paper,twocolumn,english,prb]{revtex4-1}
\usepackage[T1]{fontenc}
\usepackage[latin9]{inputenc}
\usepackage{units}
\usepackage{textcomp}
\usepackage{amsmath}
\usepackage{amssymb}
\usepackage{graphicx}
\usepackage{esint}

\makeatletter

\providecommand{\tabularnewline}{\\}

 \@ifundefined{textcolor}{}
 {%
  \definecolor{BLACK}{gray}{0}
  \definecolor{WHITE}{gray}{1}
  \definecolor{RED}{rgb}{1,0,0}
  \definecolor{GREEN}{rgb}{0,1,0}
  \definecolor{BLUE}{rgb}{0,0,1}
  \definecolor{CYAN}{cmyk}{1,0,0,0}
  \definecolor{MAGENTA}{cmyk}{0,1,0,0}
  \definecolor{YELLOW}{cmyk}{0,0,1,0}
  }

\makeatother

\usepackage{babel}

\makeatother

\usepackage{babel}
\begin{document}

\title{Spectrally-resolved femtosecond reflectivity relaxation dynamics
in undoped SDW 122-structure iron based pnictides}

\author{A. Pogrebna$^{1}$, N.Vuji\v{c}i\'c$^{1,2}$, T. Mertelj$^{1}$,
G. Cao$^{3}$, Z. A. Xu$^{3}$, J.-H. Chu$^{4}$ , I. R. Fisher$^{4}$
and D. Mihailovic$^{1}$}

\affiliation{$^{1}$Complex Matter Dept., Jozef Stefan Institute, Jamova 39, Ljubljana,
SI-1000, Ljubljana, Slovenia }

\affiliation{$^{2}$Institute of Physics, Bijeni\v{c}ka 46, HR-10000 Zagreb, Croatia}

\affiliation{$^{3}$Department of Physics, Zhejiang University, Hangzhou 310027,
People\textquoteright{}s Republic of China}

\affiliation{$^{4}$Geballe Laboratory for Advanced Materials and Department of
Applied Physics, Stanford University, Stanford, California 94305,
USA}

\date{\today}
\begin{abstract}
We systematically investigate temperature- and spectrally-dependent
optical reflectivity dynamics in AAs$_{2}$Fe$_{2}$, (A=Ba, Sr and
Eu), iron-based superconductors parent spin-density-wave (SDW) compounds.
Two different relaxation processes are identified. The behavior of
the slower process, which is strongly sensitive to the magneto-structural
transition, is analyzed in the framework of the relaxation-bottleneck
model involving magnons. The results are compared to recent time resolved
angular photoemission results (TR-ARPES) and possible alternative
assignment of the slower relaxation to the magneto-structural order
parameter relaxation is discussed.
\end{abstract}
\maketitle

\section{Introduction}

Time domain optical spectroscopy has been, among other spectroscopies,
very instrumental in elucidating the nature of the superconducting
and unusual normal states in novel superconductors and related materials
by virtue of the fact that different components in the low-energy
excitation spectrum could be distinguished by their lifetimes.\cite{DemsarPodobnik1999,KaindlWoerner2000,AverittRodriguez2001,SegreGedik2002,KusarDemsar2005,ChiaZhu2006,LiuToda2008,CaoWei2008,ChiaTalbayev2010,TorchinskyChen2010,CoslovichGiannetti2011,MerteljKabanov2009prl,StojchevskaMertelj2012}
Unfortunately, the all-optical technique lacks the momentum resolution
so assignments of the relaxation processes were indirect, based on
model-predicted temperature and fluence dependencies.\cite{KabanovDemsar99,DemsarAveritt2003,GedikBlake2004}
Despite the indirect assignments no inconsistencies with the previous
all-optical based results were found for the cuprates by the recent
laser time-resolved angular-resolved-photoemission\cite{CortesRettig2011,SmallwoodHinton2012}
experiments.

Most of the published all-optical time resolved experiments were performed
using relatively narrow-band spectrally unresolved probes. Since narrow-band
probes can be rather selective, probing only a limited subset of the
relevant low-energy electronic states, and laser TR-ARPES suffers
from surface sensitivity and limited momentum range, broadband spectrally-resolved
all-optical transient experiments are necessary to complement both
probes and elucidate information about any additional relaxation processes,
that might have been missed due to the limitations. 

In the cuprates a few such experiments have been performed\cite{GadermaierAlexandrov2010,GiannettiCilento2011,CoslovichGiannetti2013}
indicating, that there are no additional low-temperature relaxation
processes to the ones previously observed by all-optical narrow-band
probes and associated with the pseudogap and the superconducting states.\cite{CoslovichGiannetti2013}
In iron-based superconductors, to our best knowledge, no broad-band
time-resolved spectroscopy data exist, except in the THz region.\cite{KimPashkin2012}
Similarly to the cuprates however, the narrow-band all-optical time
resolved spectroscopy indicates a limited number of distinctive relaxation
components.\cite{MerteljKabanov2009prl,StojchevskaKusar2010,ChiaTalbayev2010,StojchevskaMertelj2012,TorchinskyChen2010} 

The multiband nature of iron-based pnictides offers a possibility
of additional photoexcited quasiparticle relaxation pathways that
might remain undetected by the narrow-band optical and laser TR-ARPES\cite{RettigCortes2012}
probes. In the absence of broad-band time-resolved spectroscopy data
in iron-based pnictides and in order to check about existence of additional
processes, we therefore performed a systematic spectrally-resolved
visible broad-band probe transient-reflectivity study of the SDW state
in three related undoped parent compounds: AFe$_{2}$As$_{2}$ (A-122)
with A= Ba, Sr and Eu. 

Previously it was shown,\cite{StojchevskaKusar2010,MerteljKusar2010}
that in the SDW state a single $\sim1$ ps relaxation component dominates
the near-infrared narrow-band optical response, while TR-ARPES experiments\cite{RettigCortes2012}
indicate at least two distinct relaxation processes, with the longest
relaxation time similar to the one observed in the optical response.

Our new broadband-probe results confirm previous near-infrared narrow-band
results and suggest the existence of another faster sub-200-fs relaxation
component in addition to the previously reported slower one\cite{StojchevskaKusar2010,MerteljKusar2010,RettigCortes2012}
for all three investigated compunds. The additional component is comparable
or faster than the experimental temporal resolution of $\sim200$
fs and compatible with the fastest component measured by TR-ARPES\cite{RettigCortes2012}.
The slower, previously-observed, shows, differently from TR-ARPES,
a divergent-like relaxation time at the respective magneto-structural
transition temperatures. 

The temperature dependencies of the optical-relaxation-transient amplitudes
are analyzed and discussed in the framework of the relaxation bottleneck
due to opening of the partial charge gap in the orthorhombic SDW state.
An alternative assignment of the slower component to the SDW amplitude
mode is also discussed in relation to the recent proposal for description
of the transient optical reflectivity in the charge density wave state.\cite{SchaeferKabanov2010}

\section{Experimental}

\subsection{Samples}

Single crystals of EuFe$_{2}$As$_{2}$ and SrFe$_{2}$As$_{2}$ were
grown at Zhejiang University by a flux method, similar to a previous
report\cite{JiaoTao2011}. Small Eu chunks and powders of Fe, As (Alfa
Aesar, > 99.9\%) were mixed together in the molar ratio of Eu:Fe:As
= 1:5:5 and sealed in an evacuated quartz ampoule. After heating the
mixture up to 973 K for 24 hours, the obtained precursor was thoroughly
ground before being loaded into an alumina crucible. The crucible
was then sealed by arc welding in a tube made of stainless steel under
atmosphere of argon, and then heated up to 1573 K over 10 hours in
a muffle furnace filled with argon. After holding at 1573 K for 5
hours, the furnace was cooled down to 1223 K at the rate of 5 K/h.
followed by switching off the furnace. Large crystals with size up
to 4\texttimes{}4\texttimes{}0.6 mm$^{3}$ could be harvested. 

The as-grown crystals were characterized by X-ray diffraction, which
showed good crystallinity as well as single \textquotedblleft{}122\textquotedblright{}
phase. The exact composition of the crystals was determined by energy
dispersive X-ray spectroscopy affiliated to a field-emission scanning
electron microscope (FEI Model SIRION). The measurement precision
was better than 5\% for the elements measured.

Single crystals of BaFe$_{2}$As$_{2}$ were also grown from from
a self flux at Stanford University, and characterized as described
previously.\cite{ChuAnalytis2009,StojchevskaMertelj2012}

In all three compounds the onset of the magnetic SDW ordering is concurrent
with the structural transition from tetragonal to orthorhombic symetry
at $T_{\mathrm{SDW}}=134$ K for Ba-122,\cite{ChuAnalytis2009} 190K
for Eu-122\cite{tegelRotter2008} and 203K for Sr-122.\cite{tegelRotter2008}

\subsection{Optical setup}

Measurements of probe-photon energy ($\hbar\omega_{\mathrm{pr}}$)
dependent photoinduced reflectivity $\Delta R/R$ were performed using
a standard pump-probe technique. 50 fs optical pulses from a 250-kHz
Ti:Al$_{2}$O$_{3}$ regenerative amplifier seeded with an Ti:Al$_{2}$O$_{3}$
oscillator were split in two mutually delayed parts. The pump photons
with energy $\hbar\omega_{\mathrm{P}}=3.1$ eV were derived from one
part by the standard frequency doubling in a BBO nonlinear crystal.
The probe photons were obtained from a suprecontinuum generated in
a 2.5 mm thick Al$_{2}$O$_{3}$ plate by the second part. A short-pass
Schott-glass filter was used to suppress the strong spectral density
around 1.55 eV resulting in the useful spectral density within 1.65-2.55
eV band. The polarization of the probe photons with respect to the
crystal was controlled by a broadband half waveplate and the beam
was focused on the sample by a pair of achromatic lenses. The reflected
probe beam was collimated, dispersed by a transmissive optical grating
and focused on a 48-channel silicon PIN diode array. The array was
connected to an in-house built 64-channel integrating analog to digital
converter that was synchronized to the laser and synchronously drove
an optical chopper inserted in the pump beam.

The supercontinuum pulse is chirped and about 2 ps long. The chirp
enables recovery of the temporal resolution by spectrally resolved
detection. The chirp was calibrated from measurements of the samples
at room temperature, where the relaxation dynamics is comparable\cite{StojchevskaKusar2010,StojchevskaMertelj2012}
to the temporal resolution of the setup of $\sim200-300$ fs, depending
on the probe-photon energy.

The pump and probe beams were nearly perpendicular to the cleaved
sample surface (001). The probe polarization was oriented with respect
to the crystals to obtain the maximum/minimum amplitude of $\Delta R/R$
at low temperatures. The pump beam diameters were, depending on experimental
conditions, in a 100 $\mu$m range with a smaller probe beam diameter
of $\sim50$$\mu$m.

Due to a smaller signal/noise ratio of the broadband setup with respect
to the narrowband one the pump fluences used were higher than in our
previous narrowband work,\cite{StojchevskaKusar2010,StojchevskaMertelj2012}
in the $\sim50$ $\mu$J/cm$^{2}$ range. However, linearity of the
responses with respect to both, the pump and the probe fluences was
checked to ensure that the experiments were performed in the weak
excitation regime. 

\begin{figure}[tbh]
\begin{centering}
\includegraphics[angle=-90,width=0.98\columnwidth]{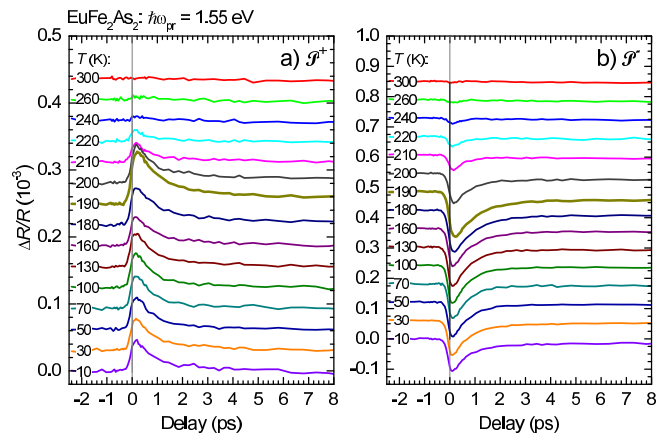} 
\par\end{centering}

\caption{(Color online) Photoinduced reflectivity transients at representative
temperatures at $\mathcal{F}\simeq10$ $\mu$J/cm$^{2}$ and 3.1 eV
pump-photon energy in EuFe$_{2}$As$_{2}$. Left and right panels
correspond to $\mathcal{P}^{+}$ and $\mathcal{P}^{-}$ polarizations,
respectively. The traces are vertically offset for clarity.}

\label{fig:figDrVsT} 
\end{figure}

\begin{figure}[tbh]
\begin{centering}
\includegraphics[angle=-90,width=0.95\columnwidth]{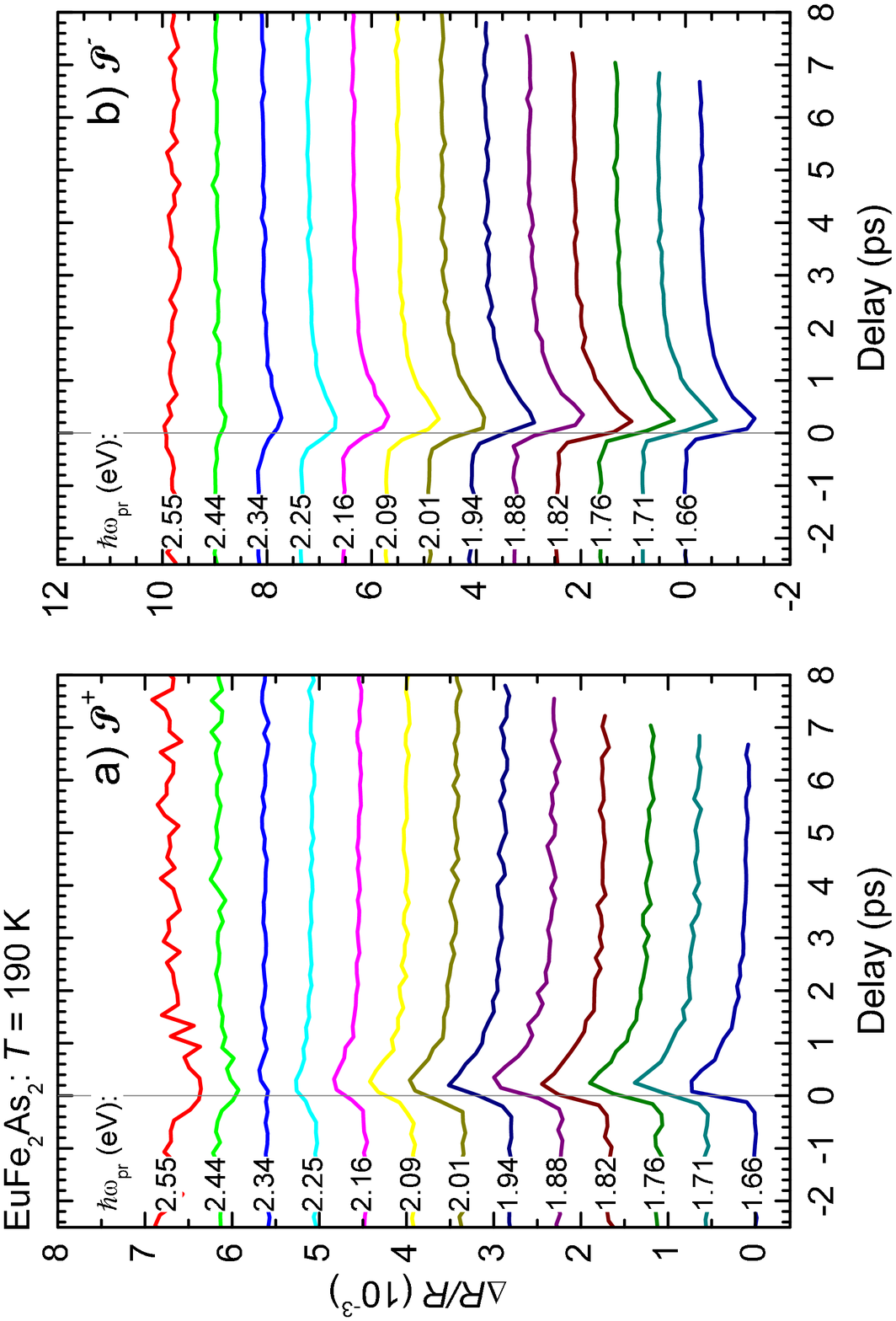}
\bigskip{}
\includegraphics[angle=-90,width=0.95\columnwidth]{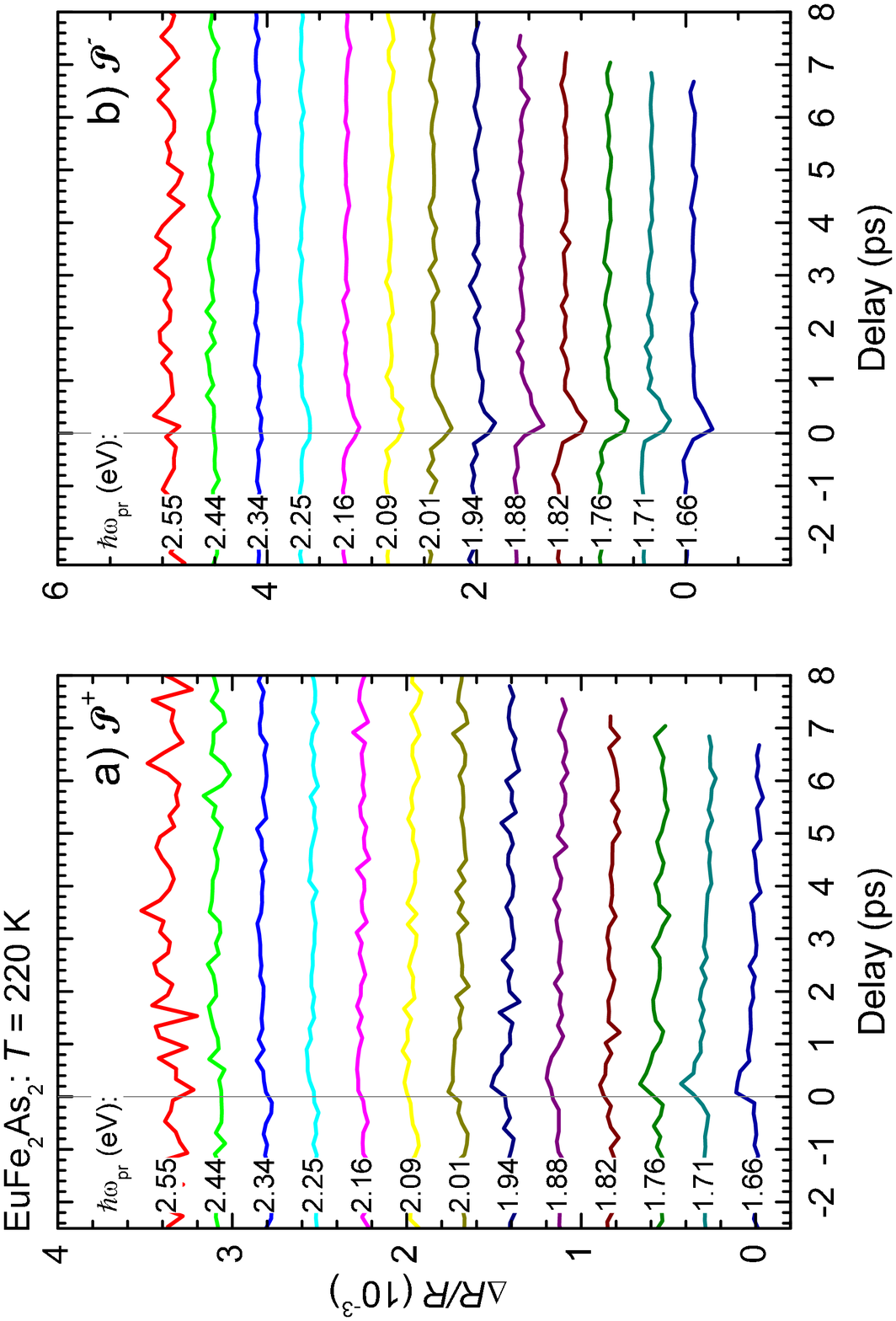}
\par\end{centering}

\caption{(Color online) Photoinduced reflectivity transients in EuFe$_{2}$As$_{2}$
as a function of the probe-photon energy at $T_{\mathrm{SDW}}$, (a)
and (b), and above $T_{\mathrm{SDW}}$, (c) and (d). (a), (c) and
(b), (d) correspond to $\mathcal{P}^{+}$ and $\mathcal{P}^{-}$ polarizations,
respectively. The traces are vertically offset for clarity.}

\label{fig-EFA-DRvsWl180K-both} 
\end{figure}

\begin{figure}[tbh]
\begin{centering}
\includegraphics[angle=-90,width=0.95\columnwidth]{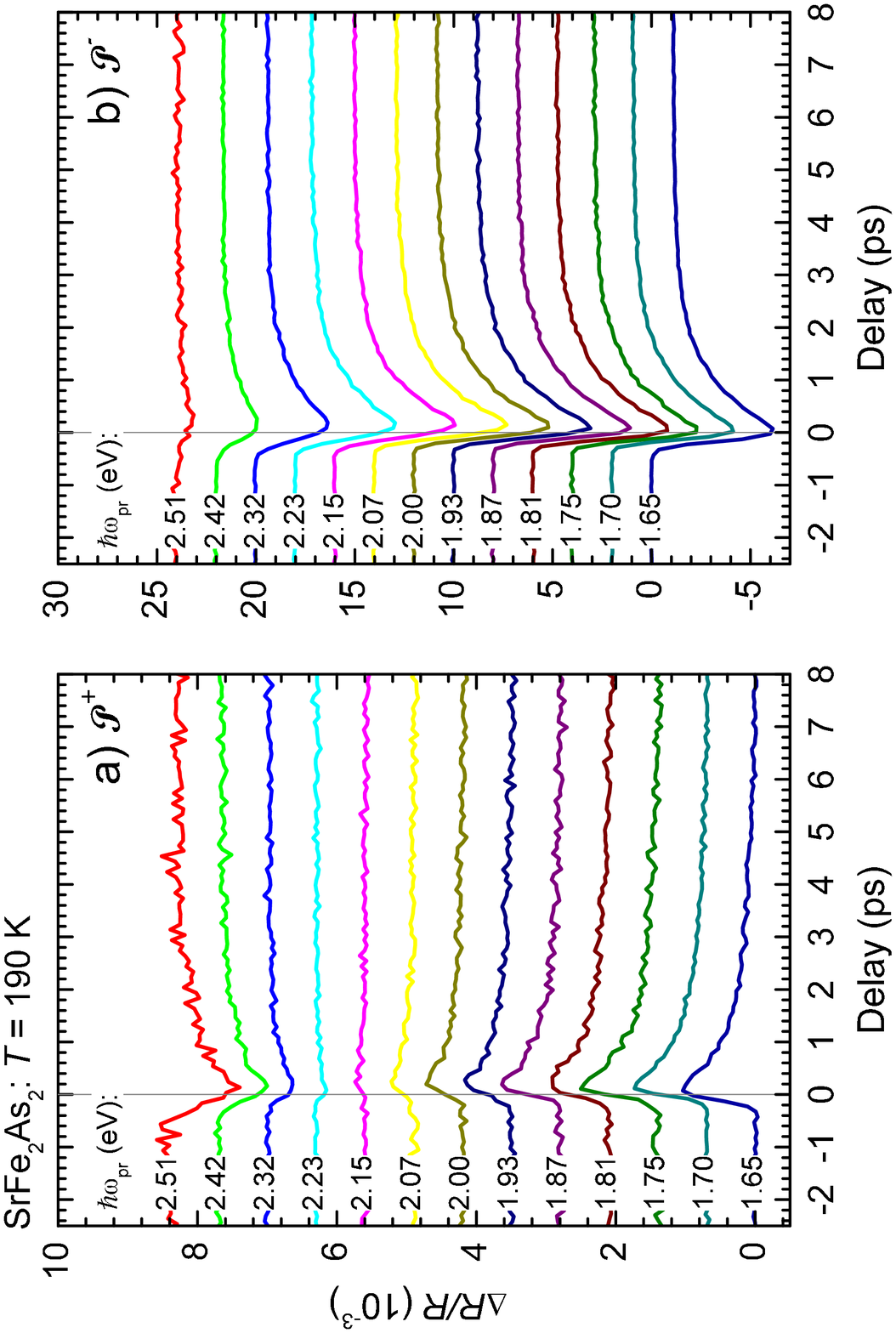}
\bigskip{}
\includegraphics[angle=-90,width=0.95\columnwidth]{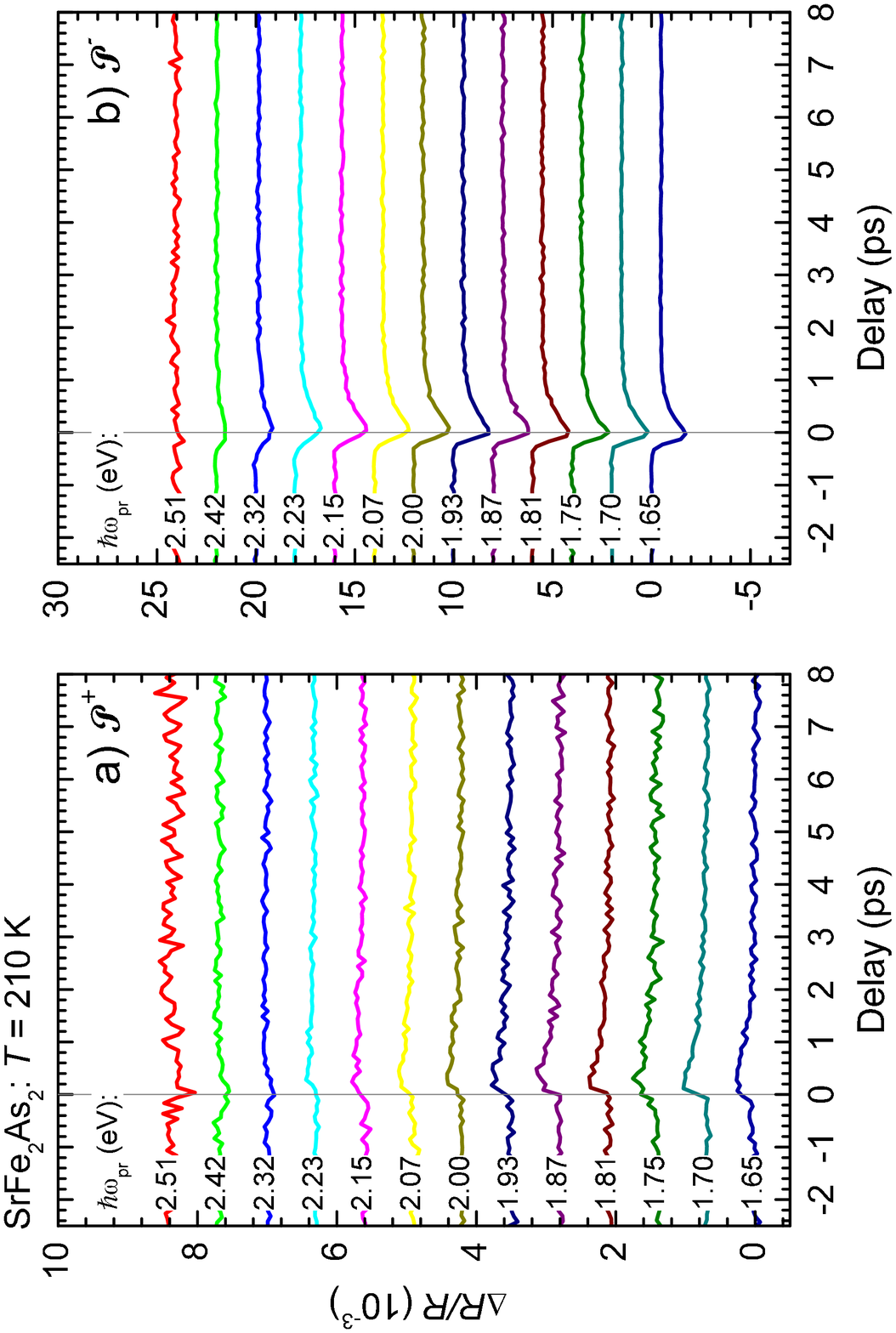}
\par\end{centering}

\caption{(Color online) Photoinduced reflectivity transients in SrFe$_{2}$As$_{2}$
as a function of the probe-photon energy just below $T_{\mathrm{SDW}}$,
(a) and (b), and above $T_{\mathrm{SDW}}$, (c) and (d). (a), (c)
and (b), (d) correspond to $\mathcal{P}^{+}$ and $\mathcal{P}^{-}$
polarizations, respectively. The traces are vertically offset for
clarity.}

\label{figSFA-DRvsWl190K-both} 
\end{figure}

\begin{figure}[tbh]
\begin{centering}
\includegraphics[angle=-90,width=0.95\columnwidth]{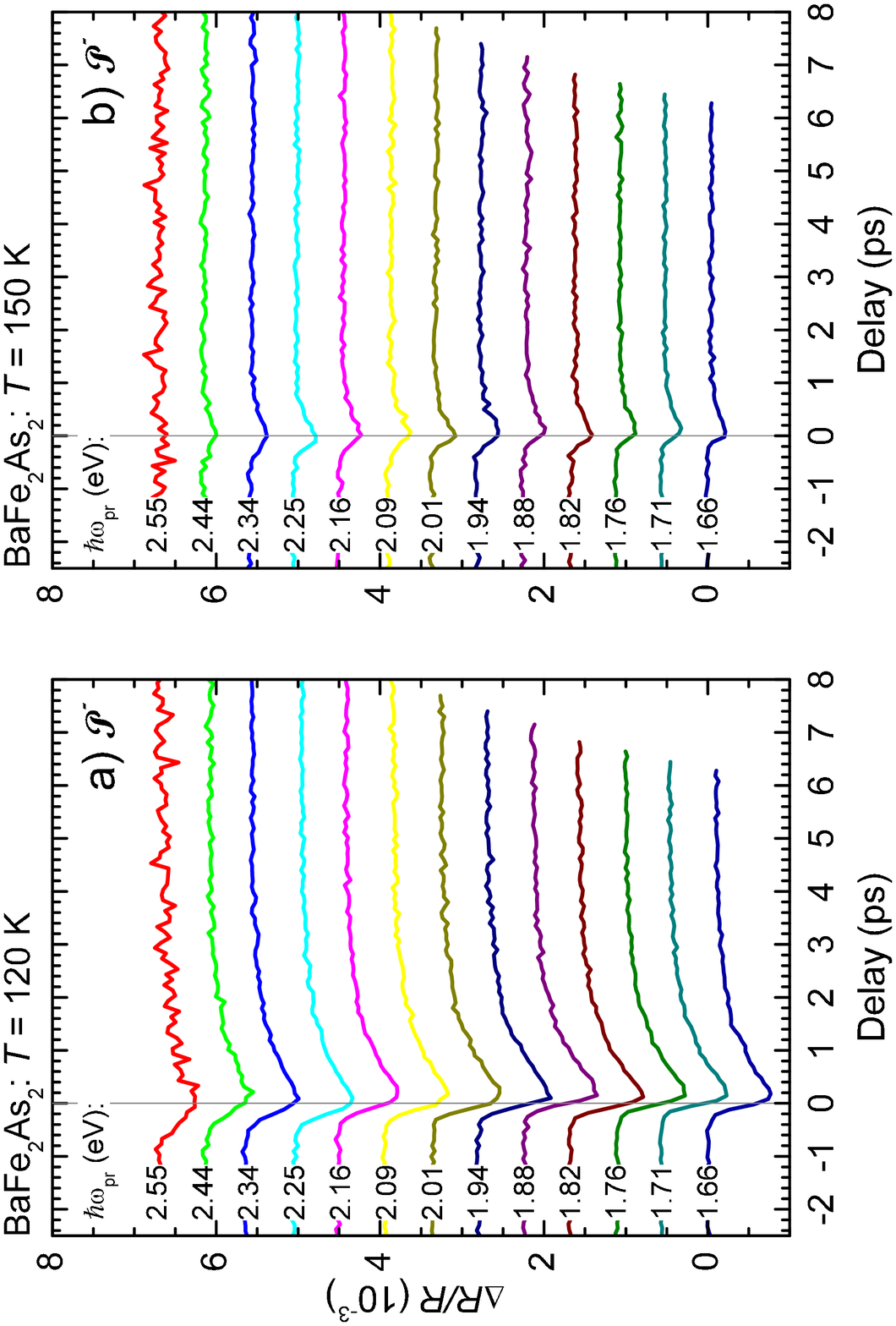} 
\par\end{centering}

\caption{(Color online) Photoinduced reflectivity transients in BaFe$_{2}$As$_{2}$
as a function of the probe-photon energy just below and just above
$T_{\mathrm{SDW}}$ (a) and (b), respectively. The traces are vertically
offset for clarity.}

\label{fig-BFA-DRvsWl-T120-T150} 
\end{figure}

\subsection{Overview of the experimental data set}

In Fig. \ref{fig:figDrVsT} we show temperature dependent $\Delta R/R$
in Eu-122 measured by narrowband 50-fs probe pulses at, $\hbar\omega_{\mathrm{pr}}=1.55$
eV, photon energy. A marked increase of $\Delta R/R$ transients amplitude
is observed around the magneto-structural transition temperature upon
cooling, consistent with previous reports in Sr-122 and Ba-122,\cite{ChiaTalbayev2010,StojchevskaKusar2010,StojchevskaMertelj2012}.
Similarly as observed in Ba-122, the transients show a 2-fold in-plane
rotational symmetry well above the tetragonal to orthorhombic magneto-structural
transition at $T_{\mathrm{SDW}}=190$ K. In Fig. \ref{fig-EFA-DRvsWl180K-both}
the probe-photon energy dependence of $\Delta R/R$ at and above $T_{\mathrm{SDW}}$
is shown for the two orthogonal polarizations. In the absence of information
about the orientation of the crystallographic axes we denote the polarizations
$\mathcal{P}^{+}$ and $\mathcal{P}^{-}$ according to the sign of
the response at the lowest probe-photon energy. With increasing $\hbar\omega_{\mathrm{pr}}$
the response for the $\mathcal{P}^{+}$ polarization changes sign
at $\hbar\omega_{\mathrm{pr}}\sim2.35$ eV. The signal for the $\mathcal{P}^{-}$
polarization, on the other hand, changes sign at a slightly larger
$\hbar\omega_{\mathrm{pr}}\sim2.5$ eV, almost at the edge of our
experimental spectral range. 

Similar behavior is observed in Sr-122 (see Fig. \ref{figSFA-DRvsWl190K-both}),
where the zero crossing is slightly lower, at $\hbar\omega_{\mathrm{pr}}\sim2.23$
eV, for the $\mathcal{P}^{+}$ polarization and slightly higher, just
at the edge of the spectral window, at $\hbar\omega_{\mathrm{pr}}\sim2.56$
eV, for the $\mathcal{P}^{-}$ polarization. 

In Ba-122 the signal is less anisotropic below $T_{\mathrm{SDW}}$\cite{StojchevskaMertelj2012}
so only the $\mathcal{P}{}^{-}$ polarization was measured, that shows
no indication of the sign change below $T\mathrm{_{SDW}}$ as shown
in Fig. \ref{fig-BFA-DRvsWl-T120-T150}.

In all three samples the spectral dependence of the transients does
not change qualitatively through the transition.

\section{Data analysis}

\begin{figure}[tbh]
\begin{centering}
\includegraphics[angle=-90,width=0.95\columnwidth]{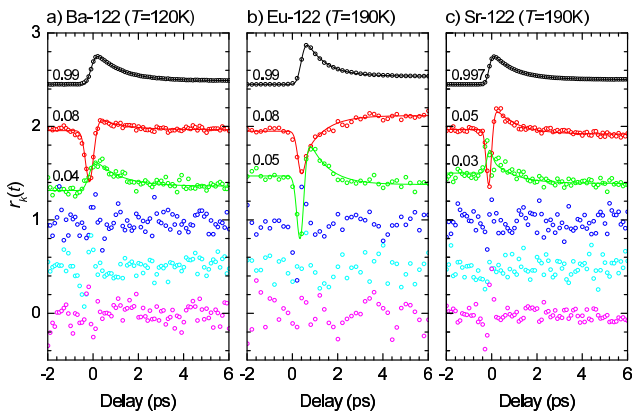} 
\par\end{centering}

\caption{(Color online) A few temporal eigenvectors with the highest singular
values. The lines are exponential fits (\ref{eq:fitfunc}) for the
three most significant temporal eigenvectors. The numerical labels
are the normalized singular values $w_{k}/\sqrt{\sum w_{l}^{2}}$.}

\label{fig-SVD-all} 
\end{figure}

\begin{figure}[tbh]
\begin{centering}
\includegraphics[angle=-90,width=0.95\columnwidth]{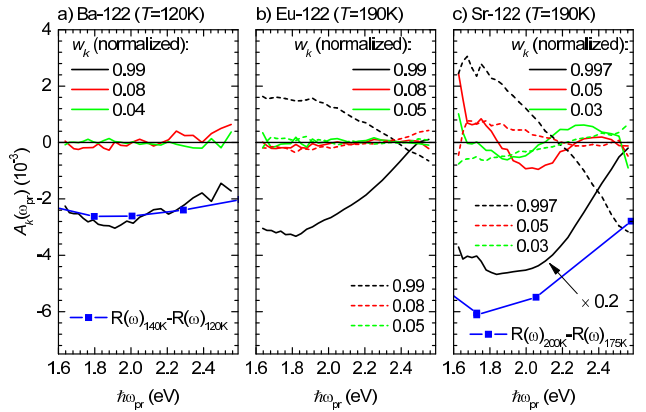} 
\par\end{centering}

\caption{(Color online) Spectral dependence of the three most significant SVD
temporal-eigenvectors weights just below respective $T_{\mathrm{SDW}}$.
The numerical labels are normalized singular values. The thermal reflectivity
difference from ref. \cite{CharnukhaLarkin2013} for Ba-122 in Sr-122
is shown for comparison.}

\label{fig-SVD-energy-all} 
\end{figure}

\begin{figure}[tbh]
\begin{centering}
\includegraphics[angle=-90,width=0.95\columnwidth]{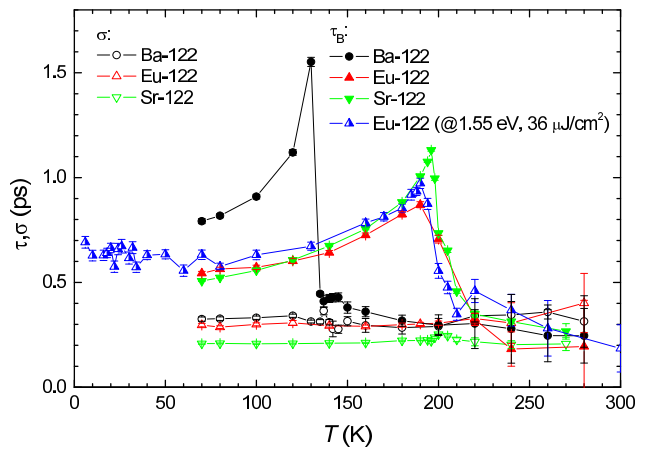} 
\par\end{centering}

\caption{(Color online) The risetime/resolution parameter and the longest relaxation
time, $\tau_{\mathrm{B}}$, as a function of $T$. The shortest relaxation
time, $\tau_{\mathrm{A}}$, is shorter than $\sigma$, so it was kept
constant at 50 fs for all fits. The 1.55-eV probe result in Eu-122
is aso shown for comparison.}

\label{fig-SVD-tau-vs-T} 
\end{figure}

\subsection{Determination of independent components}

The measured data points at each temperature can be arranged into
a rectangular matrix and decomposed using the standard singular value
decomposition (SVD) to obtain orthonormal eigenvectors describing
the data:

\begin{eqnarray}
\frac{\Delta R(\omega_{\mathrm{pr,}i},t_{j})}{R(\omega_{\mathrm{pr,}i})} & = & \sum w_{k}u_{ik}v_{jk}\nonumber \\
 & = & \sum A_{k}(\hbar\omega_{\mathrm{pr,}i})r_{k}(t_{j}).\label{eq:SVD}
\end{eqnarray}
Here $w_{k}$, $u_{ik}$ and $v_{jk}$ are the singular values (SV),
the left singular eigenvectors and the right singular eigenvectors,
respectively. With the above matrix arrangement we define the SVD
spectral weights (SW) as weighted left singular eigenvectors $A_{k}(\omega_{\mathrm{pr,}i})=w_{k}u_{ik}$,
while the right singular eigenvectors correspond to the orthonormal
temporal eigenvectors (TEv): $r_{k}(t_{j})=v_{jk}$. 

In Fig. \ref{fig-SVD-all} we plot a few most significant TEv for
each sample at temperatures just below $T_{\mathrm{SDW}}$.%
\footnote{For Sr-122 and Eu-122 we performed SVD decomposition joining together
the data for both polarizations at each $T$.%
} It is obvious, that for each sample only the three most significant
TEv contain the coherent response, while the rest represent noise
only. SW corresponding to the three most significant TEv are shown
in Fig. \ref{fig-SVD-energy-all}. The TEv with the largest SV clearly
dominates the spectral response in all compounds. In Ba-122 and Eu-122
the SW of the next two TEv are near the noise level while in Sr-122
they are almost comparable to the first TEv SW for the $\mathcal{P}^{+}$
polarization. 

The spectral dependence of the most intensive component SW is very
similar to the static reflectivity change when crossing the magneto-structural
transition\cite{CharnukhaLarkin2013} in Ba-122. In Sr-122 however,
the similarity is only in the peak position at $\sim1.7$ eV with
a large difference between the positions of the zero crossings. 

The three largest SV TEv can be rather well fit by a double exponential
relaxation model with a finite risetime/resolution and a finite long
delay value (see Fig. \ref{fig-SVD-all}):
\begin{eqnarray}
r_{k}(t) & = & \underset{m\in\{\mathrm{A,B}\}}{\sum}\frac{A_{m,k}}{2}\mathrm{e}^{-\frac{t-t_{0}}{\tau_{m}}}\operatorname{erfc}\left(\frac{\sigma^{2}-4(t-t_{0})\tau_{m}}{2\sqrt{2}\sigma\tau_{m}}\right)+\nonumber \\
 &  & +\frac{A_{\mathrm{C,k}}}{2}\operatorname{erfc}\left(-\frac{\sqrt{2}(t-t_{0})}{\sigma}\right).\label{eq:fitfunc}
\end{eqnarray}
Here the risetime/resolution parameter $\sigma$ and relaxation times
$\tau_{m}$ are shared among the three TEv, while the amplitudes $A_{l,k}$
are kept independent. 

The fits yield virtually $T$-independent $\sigma$ of 200-300 fs
shown in Fig \ref{fig-SVD-tau-vs-T} indicating that it corresponds
to the instrumental resolution.%
\footnote{Somewhat larger $\sigma$ in Ba-122 and Eu-122 might be attributed
to a slightly worse chirp compensation. %
} One of the relaxation times, $\tau_{\mathrm{A}}$, was found significantly
shorter than $\sigma$ so it was kept constant at 50 fs for all fits.
The second relaxation time $\tau_{\mathrm{B}}$ shows a marked $T$-dependence
with a divergent like behavior at $T_{SDW}$. In Ba-122 it sharply
drops above $T_{\mathrm{SDW}}$, while in Eu-122 and Sr-122 the drop
is somewhat less steep. At $T$ significantly above $T_{\mathrm{SDW}}$
it becomes comparable to the experimental temporal resolution.

\begin{figure}[tbh]
\begin{centering}
\includegraphics[angle=-90,width=0.5\columnwidth]{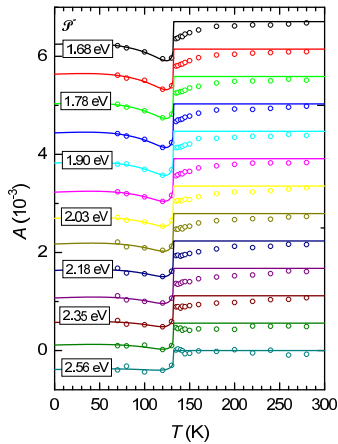} 
\par\end{centering}

\caption{(Color online) The amplitude of the $\Delta R/R$ transients in BaFe$_{2}$As$_{2}$
as a function of $T$. The thin lines are the bottleneck-model (\ref{eq:DR-npe})
fits discussed in text. The traces are vertically offset for clarity.}

\label{figAvsT-Ba} 
\end{figure}
\begin{figure}[tbh]
\begin{centering}
\includegraphics[angle=-90,width=0.95\columnwidth]{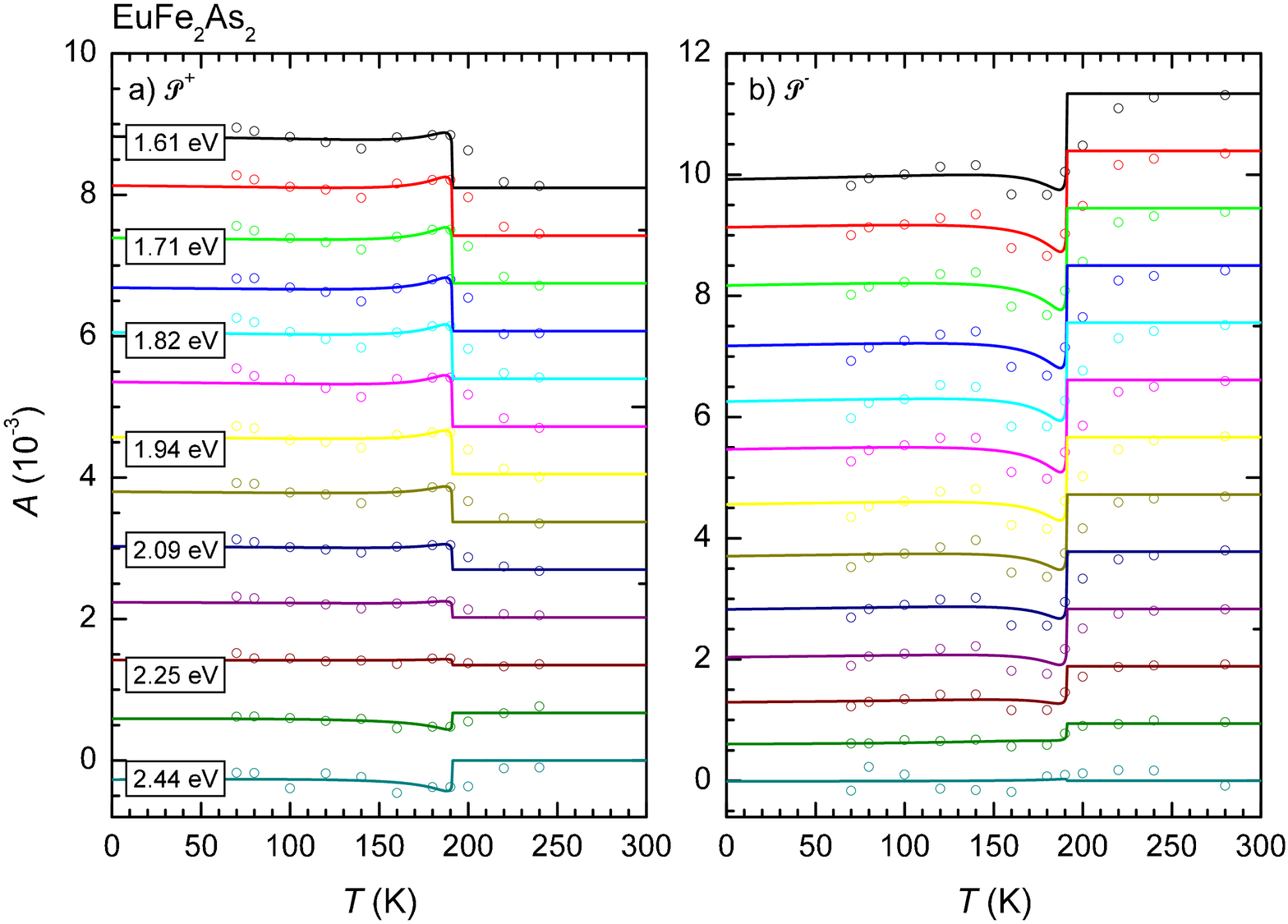} 
\par\end{centering}

\caption{(Color online) The amplitude of the $\Delta R/R$ transients in EuFe$_{2}$As$_{2}$
as a function of $T$. (a) and (b) correspond to $\mathcal{P}^{+}$
and $\mathcal{P}^{-}$ polarizations, respectively. The thin lines
are the bottleneck-model (\ref{eq:DR-npe}) fits discussed in text.
The traces are vertically offset for clarity.}

\label{figAvsT-Eu} 
\end{figure}
\begin{figure}[tbh]
\begin{centering}
\includegraphics[angle=-90,width=0.95\columnwidth]{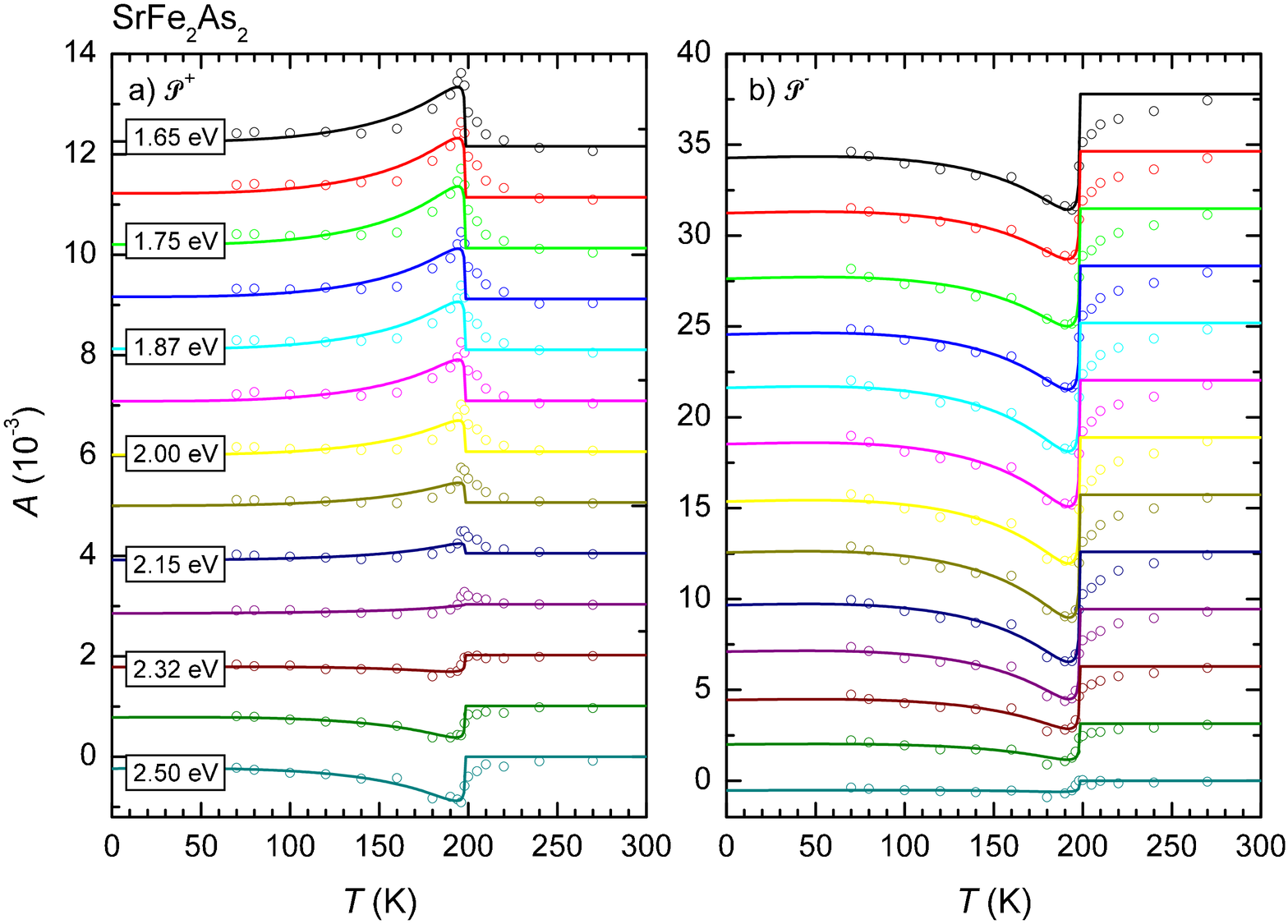} 
\par\end{centering}

\caption{(Color online) The amplitude of the $\Delta R/R$ transients in SrFe$_{2}$As$_{2}$
as a function of $T$. (a) and (b) correspond to $\mathcal{P}^{+}$
and $\mathcal{P}^{-}$ polarizations, respectively. The thin lines
are the bottleneck-model (\ref{eq:DR-npe}) fits discussed in text.
The traces are vertically offset for clarity.}

\label{figAvsT-Sr} 
\end{figure}
\begin{figure}[tbh]
\begin{centering}
\includegraphics[angle=-90,width=0.98\columnwidth]{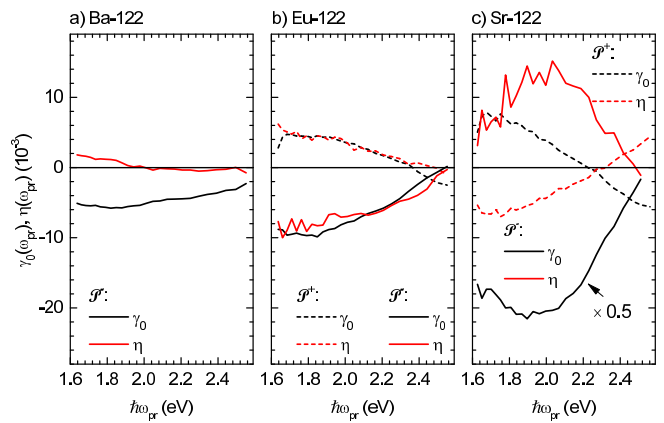} 
\par\end{centering}

\caption{(Color online) Spectral dependence of the fit amplitudes for all three
samples.}

\label{figA0-A1-vsEnergy} 
\end{figure}

\subsection{Bottleneck model fits}

To analyze the anisotropic $T$-dependence of the $\Delta R/R$ amplitude
we start with the general form of equation, that describes the photoinduced
reflectivity change due to the presence of photoexcited carriers for
a pair of bands:\cite{StojchevskaMertelj2012}
\begin{align}
\Delta R_{\alpha,\beta} & \propto\int\mathrm{d}^{3}k[\left|M_{\alpha,\beta}(\mathbf{k})\right|^{2}\Delta f_{\mathrm{\alpha}}(\mathbf{k})\times\nonumber \\
 & \times g\left(\epsilon_{\beta}(\mathbf{k})-\epsilon_{\alpha}(\mathbf{k})-\hbar\omega_{\mathrm{pr}}\right)].\label{eq:DR}
\end{align}
Here $M_{\alpha,\beta}$ is the effective probe-polarization dependent
optical-dipole matrix element between an initial band, $\alpha$,
and a final band, $\beta$, $\Delta f_{\alpha}(\mathbf{k})$ the photoexcited
change of the quasiparticle distribution function in the initial band,
$g(\epsilon)$ the effective transition line-shape and $\hbar\omega_{\mathrm{pr}}$
the probe-photon energy. For simplicity we assumed that the energy
of the final band is far from the Fermi energy, $\left|\epsilon_{\beta}(\mathbf{k})-\epsilon_{\mathrm{F}}\right|\sim\hbar\omega_{\mathrm{probe}}\gg k_{\mathrm{B}}T$,
so $\Delta f_{\beta}(\mathbf{k})$ can be neglected after the fast
initial relaxation of the ultra-hot carriers.

The integral (\ref{eq:DR}) selectively samples $\Delta f_{\alpha}(\mathbf{k})$
in different regions of the $k$-space depending on the probe polarization
and photon energy. Due to contributions of several optical transitions
with finite effective line-widths it is usually assumed that (\ref{eq:DR})
smoothly samples over the relevant energy range in the vicinity of
the Fermi energy and $\Delta R$ can be approximated by the total
photoexcited carrier density, $\Delta R=\gamma n_{\mathrm{pe}}$,\cite{KabanovDemsar99,DvorsekKabanov2002}
and any change of $\Delta R$ upon change of external parameters ($T$
for example) is attributed to the change of $n_{\mathrm{pe}}$ while
the proportionality factor $\gamma$ is assumed to be constant.

In AFe$_{2}$As$_{2}$ however, a complex band structure reorganization,
with bands shifting by as much as 80 meV, has been observed below
$T_{\mathrm{SDW}}$.\cite{YiLu2011} These shifts can significantly
modify the sampling region of the integral (\ref{eq:DR}) and violate
the assumption of a constant $\gamma$. To take this into account
we therefore assume that $\gamma$ is temperature dependent and expand
it in terms of an order parameter. The order parameter can be associated
with the opening of a partial $T$-dependent charge gap $\Delta(T)$
upon the Fermi surface reconstruction below $T\mathrm{_{SDW}}$.\cite{AnalytisMcDonald2009}
Assuming a complex BCS-like order parameter with the magnitude $\Delta(T)$
we obtain:
\begin{equation}
\mbox{\ensuremath{\Delta}}R=\left[\gamma_{0}+\eta\frac{\Delta^{2}(T)}{\Delta^{2}(0)}\right]n_{\mathrm{pe}}.\label{eq:DR-npe}
\end{equation}
To describe the $T$-dependence of $n_{\mathrm{pe}}$ we use the bottleneck
model from Kabanov \textit{et al.}\cite{KabanovDemsar99},

\begin{alignat}{1}
n_{\mathrm{pe}}\propto\nicefrac{1}{\left[\left(\frac{2\Delta\left(T\right)}{k_{\mathrm{B}}T_{\mathrm{c}}}+\frac{T}{T\mathrm{_{c}}}\right)\left(1+g_{\mathrm{ph}}\sqrt{\frac{k_{\mathrm{B}}T}{\Delta\left(T\right)}}\exp\left(-\frac{\Delta\left(T\right)}{k_{\mathrm{B}}T}\right)\right)\right],}\label{eq:npevsT}
\end{alignat}
where $g_{\mathrm{ph}}$ represents the relative effective number
of the involved boson degrees of freedom. Using the BCS temperature
dependent gap we can obtain a good fit of equation (\ref{eq:DR-npe})
to the $\Delta R/R$ amplitude for both probe polarizations (see Figs.
\ref{figAvsT-Ba}, \ref{figAvsT-Eu}, \ref{figAvsT-Sr}). The relative
gap magnitudes are consistent (see Table \ref{tbl:gaps}) with previously
reported values\cite{StojchevskaKusar2010,StojchevskaMertelj2012}
and the highest of the two gaps observed in the optical conductivity\cite{HuDong2008,CharnukhaLarkin2013},
with a larger discrepancy for Eu-122\cite{WuBarisic2009}.

\begin{table}
\medskip{}

\begin{tabular}{c|cc}
sample & $\nicefrac{2\Delta(0)}{k_{\mathrm{B}}T\mathrm{_{SDW}}}$ & $g_{\mathrm{ph}}$\tabularnewline
\hline 
Ba-122 & $8\pm6$ & $3.2\pm1.1$\tabularnewline
Eu-122 & $13\pm8$ & $2.8\pm1.4$\tabularnewline
Sr-122 & $8\pm3$ & $2.7\pm0.8$\tabularnewline
YBa$_{2}$Cu$_{3}$O$_{7-\delta}$ \cite{DemsarPodobnik1999} & $10$ & 200\tabularnewline
\end{tabular}

\caption{Charge gap magnitudes and the relative effective number of involved
bosons as obtained from the fits of (\ref{eq:DR-npe}) to the data
shown in Figs. \ref{figAvsT-Ba}, \ref{figAvsT-Eu}, \ref{figAvsT-Sr}.
The results of a fit to the data in YBa$_{2}$Cu$_{3}$O$_{7-\delta}$
superconductor ($T_{\mathrm{c}}=93$ K) is shown for comparison.}
\label{tbl:gaps}
\end{table}

\section{Discussion}

The bottleneck model (\ref{eq:npevsT}) is based on the assumption
that on a certain timescale a quasi-equilibrium is achieved between
a gaped quasiparticle population and a boson population with characteristic
energy of $\hbar\omega=2\Delta(T)$. This assumption is valid when
the energy relaxation is sufficiently slow, even if the system is
not fully gaped as is the case of the investigated SDW iron pnictides,
that remain metallic below $T_{\mathrm{SDW}}$. The quasi-equilibrium
is achieved on a timescale of $\sim200$ fs and the energy relaxation
time is in the $0.5$-$1.5$ ps range. Based on the room-$T$ relaxation
times,\cite{StojchevskaKusar2010,StojchevskaMertelj2012} where the
relaxation is governed by the electron phonon interaction,\cite{GadermaierAlexandrov2010,gadermaierKabanov2012}
the risetime is long enough for establishing a quasi-equilibrium between
the qusiparticle and high frequency phonon populations. On the other
hand, the characteristic boson energies obtained from the fits are
in the range $\sim100-\mbox{\ensuremath{\sim}}200$ meV and significantly
exceed the maximum phonon energy that is $<50$ meV.\cite{ZbiriMittal2010}
The phonon bottleneck is therefore \emph{not consistent} with the
fit results, except near the transition, where $\Delta(T)$ is smaller.

Another possible boson reservoir are antiferromagnetic magnons with
maximum energies exceeding $\sim200$ meV.\cite{EwingsPerring2008,EwingsPerring2011,SugaiMizuno2013}
The number of magnon degrees of freedom is however much smaller than
the number of phonon degrees of freedom and should be reflected in
$g_{\mathrm{ph}}$. Indeed, the obtained values of $g_{\mathrm{ph}}$
are significantly smaller than in the cuprate superconductor YBa$_{2}$Cu$_{3}$O$_{7-\delta}$
(see. Table \ref{tbl:gaps}), where the phonon bottleneck was clearly
established. 

The bottleneck model therefore consistently describes the data including
the divergent-like increase\cite{KabanovDemsar99} of $\tau_{\mathrm{B}}$
at the magnetostructural transition and the obtained gap parameters
are consistent with results obtained by means of the steady state
optical spectroscopy\cite{HuDong2008,WuBarisic2009,CharnukhaLarkin2013}
.

Comparing our data with the TR-ARPES results in Eu-122\cite{RettigCortes2012}
reveals agreement between the relaxation time of the resolution-limited
fast component observed in our case and the $T$-dependent hole relaxation
time of $\sim200-300$ fs observed by TR-ARPES. Contrary, the $T$-dependent
relaxation-time of the dominant slower component observed in our case
differs from the slower TR-ARPES electron-relaxation time beyond the
combined error bars of the experiments. While the low-$T$ optical
value of 650 fs is just slightly smaller than the TR-ARPES value of
800 fs, at $T_{\mathrm{SDW}}$ the optical relaxation time shows a
clear divergent-like behavior, that is absent in TR-ARPES data. This
indicates that either, (i) the optical probe and TR-ARPES probe detect
different relaxation processes, or (ii) there are some unidentified
surface effects that influence the relaxation near the surface probed
by TR-ARPES preventing divergence. Since more systematic TR-ARPES
investigations are needed to clarify possibility (ii) let us focus
on possibility (i).

The photoinduced dielectric constant changes captured in Eq. (\ref{eq:DR})
and taken into account in the bottleneck model are limited to the
change of the nonequilibrium quasiparticle distribution function and
neglect any change in $\epsilon(\mathbf{k}$) or the matrix elements.
In collectively ordered systems such are density waves, however, the
collective degrees of freedom can be significantly perturbed by the
photoexcitation affecting both $\epsilon(\mathbf{k}$) and matrix
elements. It was proposed recently\cite{SchaeferKabanov2010,SchaeferKabanov2014}
that the relaxation dynamics in charge-density waves can be described
by coupled dynamics of an overdamped electronic amplitude mode and
phonon modes, without directly invoking the photoinduced-quasiparticles
absorption scenario. In SDW systems a similar overdamped Raman active
amplitude mode was predicted\cite{psaltakis1984,gruner1994}, but,
to our best knowledge, never observed.

Despite the fact that the bottleneck relaxation model discussed above
describes the data below $T_{\mathrm{SDW}}$ rather consistently in
a broad frequency range, there exists a possibility that the observed
optical relaxation is due to the order parameter relaxation, the SDW
\emph{amplitude mode}. Indeed, the gap dynamics from TR-ARPES\cite{RettigCortes2012}
shows 100-200 fs risetime and a sub-picosecond decay consistent with
the slower optical relaxation timescales. Moreover, the transient
optical response shows a dominant single-component relaxation over
a broad spectral range with the spectral dependence (Fig. \ref{fig-SVD-energy-all})
similar to the equilibrium phase-transition-induced reflectivity change,\cite{CharnukhaLarkin2013}%
\footnote{ Since the transition-induced reflectivity change would be hardly
explained just by $T$-induced change of the quasiparticle distribution
function at the experimental\cite{CharnukhaLarkin2013} 15-20 K $T$
difference, it should correspond to the characteristic order-parameter-induced
spectral change.%
} suggesting the transient order-parameter modulation origin of the
optical response. The divergent behavior of the relaxation time near
$T_{\mathrm{SDW}}$ is also consistent with the critical slowing down
of the amplitude mode at the transition.

Above $T_{\mathrm{SDW}}$ the spectral signature of the response,
that remains and is clearly breaking the tetragonal symmetry, is similar
as in the SDW state. It was previously assigned\cite{StojchevskaMertelj2012,MerteljStojchevska2013}
to the bottleneck due to the normal-state pseudogap and related to
the nematic fluctuations. However, the same reasoning as above can
be applied and assign it to the fluctuations of the magneto-structural
order parameter.

The two alternative origins of the optical transient response, based
on the microscopic nonequilibrium quasiparticle distribution function
$\Delta f(\mathbf{k})$ in Eq. (\ref{eq:DR}) on one hand, and a macroscopic
order parameter on the other, are however not completely independent
nor exclusive. In a BCS-like scenario the order parameter is directly
related to the quasiparticle distribution function through the gap
equation. Provided that the condensate dynamics is faster than the
kinetics of $\Delta f(\mathbf{k})$, the time evolution of the order
parameter is direcly governed by the $\Delta f(\mathbf{k})$ kinetics.\cite{elesinKopaev1981} 

In the SDW state the characteristic frequency of the pure electronic
SDW amplitude mode is predicted\cite{psaltakis1984} to be $2\Delta/\hbar$.
Taking $2\Delta\sim200$ meV would lead to $\sim3$ fs timescale which
is much faster than the observed relaxation time. This would suggest
that, irrespective of the origin of the optical coupling, the relaxation
dynamics is governed by the bottleneck physics.

In the iron-based pnictides, however, the SDW order is clearly related
to the structural transition, so it would not be surprising if the
amplitude mode effective mass was renormalized due to coupling with
the lattice, just like in charge density wave systems. In such case,
the dynamics in the presence of a bottleneck is more complicated,
leading to at least two distinct relaxation timescales, related to
the bottleneck relaxation and the magneto-structural order parameter
relaxation. In the absence of a bottleneck, however, a single magneto-structural
order parameter relaxation timescale is expected.

Assuming that the fast component in our data corresponds to the relaxation
at the ungapped parts of the Fermi surface, as suggested by Rettig
et al.\cite{RettigCortes2012}, the presence of the single-component
exponential slower relaxation is compatible with either the pure bottleneck-driven
dynamics with a fast amplitude mode dynamics or the complete absence
of any bottleneck with slow pure magneto-structural order parmeter
dynamics. 

By taking into account the bottleneck-like amplitude $T$-dependencies
and the indication of a fast initial supression of the gap upon photoexcitation
by TR-ARPES\cite{RettigCortes2012}, the experimental data are in
favor of the pure bottleneck-driven dynamics with the fast amplitude
mode dynamics. The difference in the relaxation time behavior near
the transition between optics and TR-ARPES could then be attributed
to surface effects.

The process responsible for the modulation of the dielectric constant,
however, can be a combination of the direct photoinduced absorption
given by Eq. (\ref{eq:DR}) and the indirect coupling through the
order parameter modulation of $\epsilon(\mathbf{k}$) and matrix elements,
as suggested by the similarity of the spectral dispersions between
the transient photoinduced and the static phase-transition induced\cite{CharnukhaLarkin2013}
reflectivity changes.

\section{Summary and conclusions}

A systematic all-optical spectrally-resolved transient reflectivity
study in AFe$_{2}$As$_{2}$ is presented for the first time. Two
distinct relaxation components are identified. 

(i) A fast one, previously unobserved by the near-infrared narrow-band
probe, which is faster or similar to the experimental temporal resolution
of $\sim200$ fs, is consistent with recent TR-ARPES\cite{RettigCortes2012}
results and could be associated with relaxation at the Fermi surface. 

(ii) The slower one on the timescale of 0.6-1.5 ps, observed previously
by the near-infrared narrow-band probe, with strong sensitivity to
the magneto-structural transition, contrary to TR-ARPES, slows down
near the magneto-structural transition. It can be well quantitatively
described in a \emph{broad spectral range} by a magnon-bottleneck
model, with zero temperature gap magnitudes $2\Delta(0)/T_{\mathrm{SDW}}\sim8$,
that are consistent with steady state optical spectroscopy results.

An alternative assignment of the response to the SDW magneto-structural
order-parameter amplitude mode dynamics was also discussed, however,
the present experimental data are in favor of the pure bottleneck-driven
dynamics. 
\begin{acknowledgments}
Work at Jozef Stefan Institute was supported by ARRS (Grant No. P1-0040)
and ERC grant \emph{Trajectory}. Work at Zhejiang University was supported
by the National Science Foundation of China. We would also like to
thank V. Kabanov for comments and fruitful discussions.
\end{acknowledgments}
\bibliography{biblio}

\begin{thebibliography}{48}%
\makeatletter
\providecommand \@ifxundefined [1]{%
 \@ifx{#1\undefined}
}%
\providecommand \@ifnum [1]{%
 \ifnum #1\expandafter \@firstoftwo
 \else \expandafter \@secondoftwo
 \fi
}%
\providecommand \@ifx [1]{%
 \ifx #1\expandafter \@firstoftwo
 \else \expandafter \@secondoftwo
 \fi
}%
\providecommand \natexlab [1]{#1}%
\providecommand \enquote  [1]{``#1''}%
\providecommand \bibnamefont  [1]{#1}%
\providecommand \bibfnamefont [1]{#1}%
\providecommand \citenamefont [1]{#1}%
\providecommand \href@noop [0]{\@secondoftwo}%
\providecommand \href [0]{\begingroup \@sanitize@url \@href}%
\providecommand \@href[1]{\@@startlink{#1}\@@href}%
\providecommand \@@href[1]{\endgroup#1\@@endlink}%
\providecommand \@sanitize@url [0]{\catcode `\\12\catcode `\$12\catcode
  `\&12\catcode `\#12\catcode `\^12\catcode `\_12\catcode `\%12\relax}%
\providecommand \@@startlink[1]{}%
\providecommand \@@endlink[0]{}%
\providecommand \url  [0]{\begingroup\@sanitize@url \@url }%
\providecommand \@url [1]{\endgroup\@href {#1}{\urlprefix }}%
\providecommand \urlprefix  [0]{URL }%
\providecommand \Eprint [0]{\href }%
\providecommand \doibase [0]{http://dx.doi.org/}%
\providecommand \selectlanguage [0]{\@gobble}%
\providecommand \bibinfo  [0]{\@secondoftwo}%
\providecommand \bibfield  [0]{\@secondoftwo}%
\providecommand \translation [1]{[#1]}%
\providecommand \BibitemOpen [0]{}%
\providecommand \bibitemStop [0]{}%
\providecommand \bibitemNoStop [0]{.\EOS\space}%
\providecommand \EOS [0]{\spacefactor3000\relax}%
\providecommand \BibitemShut  [1]{\csname bibitem#1\endcsname}%
\let\auto@bib@innerbib\@empty
\bibitem [{\citenamefont {Demsar}\ \emph {et~al.}(1999)\citenamefont {Demsar},
  \citenamefont {Podobnik}, \citenamefont {Kabanov}, \citenamefont {Wolf},\
  and\ \citenamefont {Mihailovic}}]{DemsarPodobnik1999}%
  \BibitemOpen
  \bibfield  {author} {\bibinfo {author} {\bibfnamefont {J.}~\bibnamefont
  {Demsar}}, \bibinfo {author} {\bibfnamefont {B.}~\bibnamefont {Podobnik}},
  \bibinfo {author} {\bibfnamefont {V.~V.}\ \bibnamefont {Kabanov}}, \bibinfo
  {author} {\bibfnamefont {T.}~\bibnamefont {Wolf}}, \ and\ \bibinfo {author}
  {\bibfnamefont {D.}~\bibnamefont {Mihailovic}},\ }\href {\doibase
  10.1103/PhysRevLett.82.4918} {\bibfield  {journal} {\bibinfo  {journal}
  {Phys. Rev. Lett.}\ }\textbf {\bibinfo {volume} {82}},\ \bibinfo {pages}
  {4918} (\bibinfo {year} {1999})}\BibitemShut {NoStop}%
\bibitem [{\citenamefont {Kaindl}\ \emph {et~al.}(2000)\citenamefont {Kaindl},
  \citenamefont {Woerner}, \citenamefont {Elsaesser}, \citenamefont {Smith},
  \citenamefont {Ryan}, \citenamefont {Farnan}, \citenamefont {McCurry},\ and\
  \citenamefont {Walmsley}}]{KaindlWoerner2000}%
  \BibitemOpen
  \bibfield  {author} {\bibinfo {author} {\bibfnamefont {R.}~\bibnamefont
  {Kaindl}}, \bibinfo {author} {\bibfnamefont {M.}~\bibnamefont {Woerner}},
  \bibinfo {author} {\bibfnamefont {T.}~\bibnamefont {Elsaesser}}, \bibinfo
  {author} {\bibfnamefont {D.}~\bibnamefont {Smith}}, \bibinfo {author}
  {\bibfnamefont {J.}~\bibnamefont {Ryan}}, \bibinfo {author} {\bibfnamefont
  {G.}~\bibnamefont {Farnan}}, \bibinfo {author} {\bibfnamefont
  {M.}~\bibnamefont {McCurry}}, \ and\ \bibinfo {author} {\bibfnamefont
  {D.}~\bibnamefont {Walmsley}},\ }\href@noop {} {\bibfield  {journal}
  {\bibinfo  {journal} {Science}\ }\textbf {\bibinfo {volume} {287}},\ \bibinfo
  {pages} {470} (\bibinfo {year} {2000})}\BibitemShut {NoStop}%
\bibitem [{\citenamefont {Averitt}\ \emph {et~al.}(2001)\citenamefont
  {Averitt}, \citenamefont {Rodriguez}, \citenamefont {Lobad}, \citenamefont
  {Siders}, \citenamefont {Trugman},\ and\ \citenamefont
  {Taylor}}]{AverittRodriguez2001}%
  \BibitemOpen
  \bibfield  {author} {\bibinfo {author} {\bibfnamefont {R.~D.}\ \bibnamefont
  {Averitt}}, \bibinfo {author} {\bibfnamefont {G.}~\bibnamefont {Rodriguez}},
  \bibinfo {author} {\bibfnamefont {A.~I.}\ \bibnamefont {Lobad}}, \bibinfo
  {author} {\bibfnamefont {J.~L.~W.}\ \bibnamefont {Siders}}, \bibinfo {author}
  {\bibfnamefont {S.~A.}\ \bibnamefont {Trugman}}, \ and\ \bibinfo {author}
  {\bibfnamefont {A.~J.}\ \bibnamefont {Taylor}},\ }\href {\doibase
  10.1103/PhysRevB.63.140502} {\bibfield  {journal} {\bibinfo  {journal} {Phys.
  Rev. B}\ }\textbf {\bibinfo {volume} {63}},\ \bibinfo {pages} {140502}
  (\bibinfo {year} {2001})}\BibitemShut {NoStop}%
\bibitem [{\citenamefont {Segre}\ \emph {et~al.}(2002)\citenamefont {Segre},
  \citenamefont {Gedik}, \citenamefont {Orenstein}, \citenamefont {Bonn},
  \citenamefont {Liang},\ and\ \citenamefont {Hardy}}]{SegreGedik2002}%
  \BibitemOpen
  \bibfield  {author} {\bibinfo {author} {\bibfnamefont {G.~P.}\ \bibnamefont
  {Segre}}, \bibinfo {author} {\bibfnamefont {N.}~\bibnamefont {Gedik}},
  \bibinfo {author} {\bibfnamefont {J.}~\bibnamefont {Orenstein}}, \bibinfo
  {author} {\bibfnamefont {D.~A.}\ \bibnamefont {Bonn}}, \bibinfo {author}
  {\bibfnamefont {R.}~\bibnamefont {Liang}}, \ and\ \bibinfo {author}
  {\bibfnamefont {W.~N.}\ \bibnamefont {Hardy}},\ }\href {\doibase
  10.1103/PhysRevLett.88.137001} {\bibfield  {journal} {\bibinfo  {journal}
  {Phys. Rev. Lett.}\ }\textbf {\bibinfo {volume} {88}},\ \bibinfo {pages}
  {137001} (\bibinfo {year} {2002})}\BibitemShut {NoStop}%
\bibitem [{\citenamefont {Kusar}\ \emph {et~al.}(2005)\citenamefont {Kusar},
  \citenamefont {Demsar}, \citenamefont {Mihailovic},\ and\ \citenamefont
  {Sugai}}]{KusarDemsar2005}%
  \BibitemOpen
  \bibfield  {author} {\bibinfo {author} {\bibfnamefont {P.}~\bibnamefont
  {Kusar}}, \bibinfo {author} {\bibfnamefont {J.}~\bibnamefont {Demsar}},
  \bibinfo {author} {\bibfnamefont {D.}~\bibnamefont {Mihailovic}}, \ and\
  \bibinfo {author} {\bibfnamefont {S.}~\bibnamefont {Sugai}},\ }\href
  {\doibase 10.1103/PhysRevB.72.014544} {\bibfield  {journal} {\bibinfo
  {journal} {Phys. Rev. B}\ }\textbf {\bibinfo {volume} {72}},\ \bibinfo
  {pages} {014544} (\bibinfo {year} {2005})}\BibitemShut {NoStop}%
\bibitem [{\citenamefont {Chia}\ \emph {et~al.}(2006)\citenamefont {Chia},
  \citenamefont {Zhu}, \citenamefont {Lee}, \citenamefont {Hur}, \citenamefont
  {Moreno}, \citenamefont {Bauer}, \citenamefont {Durakiewicz}, \citenamefont
  {Averitt}, \citenamefont {Sarrao},\ and\ \citenamefont
  {Taylor}}]{ChiaZhu2006}%
  \BibitemOpen
  \bibfield  {author} {\bibinfo {author} {\bibfnamefont {E.~E.~M.}\
  \bibnamefont {Chia}}, \bibinfo {author} {\bibfnamefont {J.-X.}\ \bibnamefont
  {Zhu}}, \bibinfo {author} {\bibfnamefont {H.~J.}\ \bibnamefont {Lee}},
  \bibinfo {author} {\bibfnamefont {N.}~\bibnamefont {Hur}}, \bibinfo {author}
  {\bibfnamefont {N.~O.}\ \bibnamefont {Moreno}}, \bibinfo {author}
  {\bibfnamefont {E.~D.}\ \bibnamefont {Bauer}}, \bibinfo {author}
  {\bibfnamefont {T.}~\bibnamefont {Durakiewicz}}, \bibinfo {author}
  {\bibfnamefont {R.~D.}\ \bibnamefont {Averitt}}, \bibinfo {author}
  {\bibfnamefont {J.~L.}\ \bibnamefont {Sarrao}}, \ and\ \bibinfo {author}
  {\bibfnamefont {A.~J.}\ \bibnamefont {Taylor}},\ }\href {\doibase
  10.1103/PhysRevB.74.140409} {\bibfield  {journal} {\bibinfo  {journal} {Phys.
  Rev. B}\ }\textbf {\bibinfo {volume} {74}},\ \bibinfo {pages} {140409}
  (\bibinfo {year} {2006})}\BibitemShut {NoStop}%
\bibitem [{\citenamefont {Liu}\ \emph {et~al.}(2008)\citenamefont {Liu},
  \citenamefont {Toda}, \citenamefont {Shimatake}, \citenamefont {Momono},
  \citenamefont {Oda},\ and\ \citenamefont {Ido}}]{LiuToda2008}%
  \BibitemOpen
  \bibfield  {author} {\bibinfo {author} {\bibfnamefont {Y.~H.}\ \bibnamefont
  {Liu}}, \bibinfo {author} {\bibfnamefont {Y.}~\bibnamefont {Toda}}, \bibinfo
  {author} {\bibfnamefont {K.}~\bibnamefont {Shimatake}}, \bibinfo {author}
  {\bibfnamefont {N.}~\bibnamefont {Momono}}, \bibinfo {author} {\bibfnamefont
  {M.}~\bibnamefont {Oda}}, \ and\ \bibinfo {author} {\bibfnamefont
  {M.}~\bibnamefont {Ido}},\ }\href {\doibase 10.1103/PhysRevLett.101.137003}
  {\bibfield  {journal} {\bibinfo  {journal} {Phys. Rev. Lett.}\ }\textbf
  {\bibinfo {volume} {101}},\ \bibinfo {pages} {137003} (\bibinfo {year}
  {2008})}\BibitemShut {NoStop}%
\bibitem [{\citenamefont {Ning}\ \emph {et~al.}(2008)\citenamefont {Ning},
  \citenamefont {Yan-Feng}, \citenamefont {Ji-Min}, \citenamefont {Shi-Ping},
  \citenamefont {Qian-Sheng}, \citenamefont {Zhi-Guo},\ and\ \citenamefont
  {Pan-Ming}}]{CaoWei2008}%
  \BibitemOpen
  \bibfield  {author} {\bibinfo {author} {\bibfnamefont {C.}~\bibnamefont
  {Ning}}, \bibinfo {author} {\bibfnamefont {W.}~\bibnamefont {Yan-Feng}},
  \bibinfo {author} {\bibfnamefont {Z.}~\bibnamefont {Ji-Min}}, \bibinfo
  {author} {\bibfnamefont {Z.}~\bibnamefont {Shi-Ping}}, \bibinfo {author}
  {\bibfnamefont {Y.}~\bibnamefont {Qian-Sheng}}, \bibinfo {author}
  {\bibfnamefont {Z.}~\bibnamefont {Zhi-Guo}}, \ and\ \bibinfo {author}
  {\bibfnamefont {F.}~\bibnamefont {Pan-Ming}},\ }\href
  {http://stacks.iop.org/0256-307X/25/i=6/a=092} {\bibfield  {journal}
  {\bibinfo  {journal} {Chinese Physics Letters}\ }\textbf {\bibinfo {volume}
  {25}},\ \bibinfo {pages} {2257} (\bibinfo {year} {2008})}\BibitemShut
  {NoStop}%
\bibitem [{\citenamefont {Chia}\ \emph {et~al.}(2010)\citenamefont {Chia},
  \citenamefont {Talbayev}, \citenamefont {Zhu}, \citenamefont {Yuan},
  \citenamefont {Park}, \citenamefont {Thompson}, \citenamefont {Panagopoulos},
  \citenamefont {Chen}, \citenamefont {Luo}, \citenamefont {Wang},\ and\
  \citenamefont {Taylor}}]{ChiaTalbayev2010}%
  \BibitemOpen
  \bibfield  {author} {\bibinfo {author} {\bibfnamefont {E.~E.~M.}\
  \bibnamefont {Chia}}, \bibinfo {author} {\bibfnamefont {D.}~\bibnamefont
  {Talbayev}}, \bibinfo {author} {\bibfnamefont {J.-X.}\ \bibnamefont {Zhu}},
  \bibinfo {author} {\bibfnamefont {H.~Q.}\ \bibnamefont {Yuan}}, \bibinfo
  {author} {\bibfnamefont {T.}~\bibnamefont {Park}}, \bibinfo {author}
  {\bibfnamefont {J.~D.}\ \bibnamefont {Thompson}}, \bibinfo {author}
  {\bibfnamefont {C.}~\bibnamefont {Panagopoulos}}, \bibinfo {author}
  {\bibfnamefont {G.~F.}\ \bibnamefont {Chen}}, \bibinfo {author}
  {\bibfnamefont {J.~L.}\ \bibnamefont {Luo}}, \bibinfo {author} {\bibfnamefont
  {N.~L.}\ \bibnamefont {Wang}}, \ and\ \bibinfo {author} {\bibfnamefont
  {A.~J.}\ \bibnamefont {Taylor}},\ }\href {\doibase
  10.1103/PhysRevLett.104.027003} {\bibfield  {journal} {\bibinfo  {journal}
  {Phys. Rev. Lett.}\ }\textbf {\bibinfo {volume} {104}},\ \bibinfo {pages}
  {027003} (\bibinfo {year} {2010})}\BibitemShut {NoStop}%
\bibitem [{\citenamefont {Torchinsky}\ \emph {et~al.}(2010)\citenamefont
  {Torchinsky}, \citenamefont {Chen}, \citenamefont {Luo}, \citenamefont
  {Wang},\ and\ \citenamefont {Gedik}}]{TorchinskyChen2010}%
  \BibitemOpen
  \bibfield  {author} {\bibinfo {author} {\bibfnamefont {D.~H.}\ \bibnamefont
  {Torchinsky}}, \bibinfo {author} {\bibfnamefont {G.~F.}\ \bibnamefont
  {Chen}}, \bibinfo {author} {\bibfnamefont {J.~L.}\ \bibnamefont {Luo}},
  \bibinfo {author} {\bibfnamefont {N.~L.}\ \bibnamefont {Wang}}, \ and\
  \bibinfo {author} {\bibfnamefont {N.}~\bibnamefont {Gedik}},\ }\href
  {\doibase 10.1103/PhysRevLett.105.027005} {\bibfield  {journal} {\bibinfo
  {journal} {Phys. Rev. Lett.}\ }\textbf {\bibinfo {volume} {105}},\ \bibinfo
  {pages} {027005} (\bibinfo {year} {2010})}\BibitemShut {NoStop}%
\bibitem [{\citenamefont {Coslovich}\ \emph {et~al.}(2011)\citenamefont
  {Coslovich}, \citenamefont {Giannetti}, \citenamefont {Cilento},
  \citenamefont {Dal~Conte}, \citenamefont {Ferrini}, \citenamefont
  {Galinetto}, \citenamefont {Greven}, \citenamefont {Eisaki}, \citenamefont
  {Raichle}, \citenamefont {Liang}, \citenamefont {Damascelli},\ and\
  \citenamefont {Parmigiani}}]{CoslovichGiannetti2011}%
  \BibitemOpen
  \bibfield  {author} {\bibinfo {author} {\bibfnamefont {G.}~\bibnamefont
  {Coslovich}}, \bibinfo {author} {\bibfnamefont {C.}~\bibnamefont
  {Giannetti}}, \bibinfo {author} {\bibfnamefont {F.}~\bibnamefont {Cilento}},
  \bibinfo {author} {\bibfnamefont {S.}~\bibnamefont {Dal~Conte}}, \bibinfo
  {author} {\bibfnamefont {G.}~\bibnamefont {Ferrini}}, \bibinfo {author}
  {\bibfnamefont {P.}~\bibnamefont {Galinetto}}, \bibinfo {author}
  {\bibfnamefont {M.}~\bibnamefont {Greven}}, \bibinfo {author} {\bibfnamefont
  {H.}~\bibnamefont {Eisaki}}, \bibinfo {author} {\bibfnamefont
  {M.}~\bibnamefont {Raichle}}, \bibinfo {author} {\bibfnamefont
  {R.}~\bibnamefont {Liang}}, \bibinfo {author} {\bibfnamefont
  {A.}~\bibnamefont {Damascelli}}, \ and\ \bibinfo {author} {\bibfnamefont
  {F.}~\bibnamefont {Parmigiani}},\ }\href {\doibase
  10.1103/PhysRevB.83.064519} {\bibfield  {journal} {\bibinfo  {journal} {Phys.
  Rev. B}\ }\textbf {\bibinfo {volume} {83}},\ \bibinfo {pages} {064519}
  (\bibinfo {year} {2011})}\BibitemShut {NoStop}%
\bibitem [{\citenamefont {Mertelj}\ \emph {et~al.}(2009)\citenamefont
  {Mertelj}, \citenamefont {Kabanov}, \citenamefont {Gadermaier}, \citenamefont
  {Zhigadlo}, \citenamefont {Katrych}, \citenamefont {Karpinski},\ and\
  \citenamefont {Mihailovic}}]{MerteljKabanov2009prl}%
  \BibitemOpen
  \bibfield  {author} {\bibinfo {author} {\bibfnamefont {T.}~\bibnamefont
  {Mertelj}}, \bibinfo {author} {\bibfnamefont {V.}~\bibnamefont {Kabanov}},
  \bibinfo {author} {\bibfnamefont {C.}~\bibnamefont {Gadermaier}}, \bibinfo
  {author} {\bibfnamefont {N.}~\bibnamefont {Zhigadlo}}, \bibinfo {author}
  {\bibfnamefont {S.}~\bibnamefont {Katrych}}, \bibinfo {author} {\bibfnamefont
  {J.}~\bibnamefont {Karpinski}}, \ and\ \bibinfo {author} {\bibfnamefont
  {D.}~\bibnamefont {Mihailovic}},\ }\href@noop {} {\bibfield  {journal}
  {\bibinfo  {journal} {Physical Review Letters}\ }\textbf {\bibinfo {volume}
  {102}},\ \bibinfo {pages} {117002} (\bibinfo {year} {2009})}\BibitemShut
  {NoStop}%
\bibitem [{\citenamefont {Stojchevska}\ \emph {et~al.}(2012)\citenamefont
  {Stojchevska}, \citenamefont {Mertelj}, \citenamefont {Chu}, \citenamefont
  {Fisher},\ and\ \citenamefont {Mihailovic}}]{StojchevskaMertelj2012}%
  \BibitemOpen
  \bibfield  {author} {\bibinfo {author} {\bibfnamefont {L.}~\bibnamefont
  {Stojchevska}}, \bibinfo {author} {\bibfnamefont {T.}~\bibnamefont
  {Mertelj}}, \bibinfo {author} {\bibfnamefont {J.-H.}\ \bibnamefont {Chu}},
  \bibinfo {author} {\bibfnamefont {I.~R.}\ \bibnamefont {Fisher}}, \ and\
  \bibinfo {author} {\bibfnamefont {D.}~\bibnamefont {Mihailovic}},\
  }\href@noop {} {\bibfield  {journal} {\bibinfo  {journal} {Physical Review
  B}\ }\textbf {\bibinfo {volume} {86}},\ \bibinfo {pages} {024519} (\bibinfo
  {year} {2012})}\BibitemShut {NoStop}%
\bibitem [{\citenamefont {Kabanov}\ \emph {et~al.}(1999)\citenamefont
  {Kabanov}, \citenamefont {Demsar}, \citenamefont {Podobnik},\ and\
  \citenamefont {Mihailovic}}]{KabanovDemsar99}%
  \BibitemOpen
  \bibfield  {author} {\bibinfo {author} {\bibfnamefont {V.~V.}\ \bibnamefont
  {Kabanov}}, \bibinfo {author} {\bibfnamefont {J.}~\bibnamefont {Demsar}},
  \bibinfo {author} {\bibfnamefont {B.}~\bibnamefont {Podobnik}}, \ and\
  \bibinfo {author} {\bibfnamefont {D.}~\bibnamefont {Mihailovic}},\ }\href
  {\doibase 10.1103/PhysRevB.59.1497} {\bibfield  {journal} {\bibinfo
  {journal} {Phys. Rev. B}\ }\textbf {\bibinfo {volume} {59}},\ \bibinfo
  {pages} {1497} (\bibinfo {year} {1999})}\BibitemShut {NoStop}%
\bibitem [{\citenamefont {Demsar}\ \emph {et~al.}(2003)\citenamefont {Demsar},
  \citenamefont {Averitt}, \citenamefont {Taylor}, \citenamefont {Kabanov},
  \citenamefont {Kang}, \citenamefont {Kim}, \citenamefont {Choi},\ and\
  \citenamefont {Lee}}]{DemsarAveritt2003}%
  \BibitemOpen
  \bibfield  {author} {\bibinfo {author} {\bibfnamefont {J.}~\bibnamefont
  {Demsar}}, \bibinfo {author} {\bibfnamefont {R.~D.}\ \bibnamefont {Averitt}},
  \bibinfo {author} {\bibfnamefont {A.~J.}\ \bibnamefont {Taylor}}, \bibinfo
  {author} {\bibfnamefont {V.~V.}\ \bibnamefont {Kabanov}}, \bibinfo {author}
  {\bibfnamefont {W.~N.}\ \bibnamefont {Kang}}, \bibinfo {author}
  {\bibfnamefont {H.~J.}\ \bibnamefont {Kim}}, \bibinfo {author} {\bibfnamefont
  {E.~M.}\ \bibnamefont {Choi}}, \ and\ \bibinfo {author} {\bibfnamefont
  {S.~I.}\ \bibnamefont {Lee}},\ }\href {\doibase
  10.1103/PhysRevLett.91.267002} {\bibfield  {journal} {\bibinfo  {journal}
  {Phys. Rev. Lett.}\ }\textbf {\bibinfo {volume} {91}},\ \bibinfo {pages}
  {267002} (\bibinfo {year} {2003})}\BibitemShut {NoStop}%
\bibitem [{\citenamefont {Gedik}\ \emph {et~al.}(2004)\citenamefont {Gedik},
  \citenamefont {Blake}, \citenamefont {Spitzer}, \citenamefont {Orenstein},
  \citenamefont {Liang}, \citenamefont {Bonn},\ and\ \citenamefont
  {Hardy}}]{GedikBlake2004}%
  \BibitemOpen
  \bibfield  {author} {\bibinfo {author} {\bibfnamefont {N.}~\bibnamefont
  {Gedik}}, \bibinfo {author} {\bibfnamefont {P.}~\bibnamefont {Blake}},
  \bibinfo {author} {\bibfnamefont {R.~C.}\ \bibnamefont {Spitzer}}, \bibinfo
  {author} {\bibfnamefont {J.}~\bibnamefont {Orenstein}}, \bibinfo {author}
  {\bibfnamefont {R.}~\bibnamefont {Liang}}, \bibinfo {author} {\bibfnamefont
  {D.~A.}\ \bibnamefont {Bonn}}, \ and\ \bibinfo {author} {\bibfnamefont
  {W.~N.}\ \bibnamefont {Hardy}},\ }\href {\doibase 10.1103/PhysRevB.70.014504}
  {\bibfield  {journal} {\bibinfo  {journal} {Phys. Rev. B}\ }\textbf {\bibinfo
  {volume} {70}},\ \bibinfo {pages} {014504} (\bibinfo {year}
  {2004})}\BibitemShut {NoStop}%
\bibitem [{\citenamefont {Cort\'es}\ \emph {et~al.}(2011)\citenamefont
  {Cort\'es}, \citenamefont {Rettig}, \citenamefont {Yoshida}, \citenamefont
  {Eisaki}, \citenamefont {Wolf},\ and\ \citenamefont
  {Bovensiepen}}]{CortesRettig2011}%
  \BibitemOpen
  \bibfield  {author} {\bibinfo {author} {\bibfnamefont {R.}~\bibnamefont
  {Cort\'es}}, \bibinfo {author} {\bibfnamefont {L.}~\bibnamefont {Rettig}},
  \bibinfo {author} {\bibfnamefont {Y.}~\bibnamefont {Yoshida}}, \bibinfo
  {author} {\bibfnamefont {H.}~\bibnamefont {Eisaki}}, \bibinfo {author}
  {\bibfnamefont {M.}~\bibnamefont {Wolf}}, \ and\ \bibinfo {author}
  {\bibfnamefont {U.}~\bibnamefont {Bovensiepen}},\ }\href {\doibase
  10.1103/PhysRevLett.107.097002} {\bibfield  {journal} {\bibinfo  {journal}
  {Phys. Rev. Lett.}\ }\textbf {\bibinfo {volume} {107}},\ \bibinfo {pages}
  {097002} (\bibinfo {year} {2011})}\BibitemShut {NoStop}%
\bibitem [{\citenamefont {Smallwood}\ \emph {et~al.}(2012)\citenamefont
  {Smallwood}, \citenamefont {Hinton}, \citenamefont {Jozwiak}, \citenamefont
  {Zhang}, \citenamefont {Koralek}, \citenamefont {Eisaki}, \citenamefont
  {Lee}, \citenamefont {Orenstein},\ and\ \citenamefont
  {Lanzara}}]{SmallwoodHinton2012}%
  \BibitemOpen
  \bibfield  {author} {\bibinfo {author} {\bibfnamefont {C.~L.}\ \bibnamefont
  {Smallwood}}, \bibinfo {author} {\bibfnamefont {J.~P.}\ \bibnamefont
  {Hinton}}, \bibinfo {author} {\bibfnamefont {C.}~\bibnamefont {Jozwiak}},
  \bibinfo {author} {\bibfnamefont {W.}~\bibnamefont {Zhang}}, \bibinfo
  {author} {\bibfnamefont {J.~D.}\ \bibnamefont {Koralek}}, \bibinfo {author}
  {\bibfnamefont {H.}~\bibnamefont {Eisaki}}, \bibinfo {author} {\bibfnamefont
  {D.-H.}\ \bibnamefont {Lee}}, \bibinfo {author} {\bibfnamefont
  {J.}~\bibnamefont {Orenstein}}, \ and\ \bibinfo {author} {\bibfnamefont
  {A.}~\bibnamefont {Lanzara}},\ }\href@noop {} {\bibfield  {journal} {\bibinfo
   {journal} {Science}\ }\textbf {\bibinfo {volume} {336}},\ \bibinfo {pages}
  {1137} (\bibinfo {year} {2012})}\BibitemShut {NoStop}%
\bibitem [{\citenamefont {Gadermaier}\ \emph {et~al.}(2010)\citenamefont
  {Gadermaier}, \citenamefont {Alexandrov}, \citenamefont {Kabanov},
  \citenamefont {Kusar}, \citenamefont {Mertelj}, \citenamefont {Yao},
  \citenamefont {Manzoni}, \citenamefont {Brida}, \citenamefont {Cerullo},\
  and\ \citenamefont {Mihailovic}}]{GadermaierAlexandrov2010}%
  \BibitemOpen
  \bibfield  {author} {\bibinfo {author} {\bibfnamefont {C.}~\bibnamefont
  {Gadermaier}}, \bibinfo {author} {\bibfnamefont {A.~S.}\ \bibnamefont
  {Alexandrov}}, \bibinfo {author} {\bibfnamefont {V.~V.}\ \bibnamefont
  {Kabanov}}, \bibinfo {author} {\bibfnamefont {P.}~\bibnamefont {Kusar}},
  \bibinfo {author} {\bibfnamefont {T.}~\bibnamefont {Mertelj}}, \bibinfo
  {author} {\bibfnamefont {X.}~\bibnamefont {Yao}}, \bibinfo {author}
  {\bibfnamefont {C.}~\bibnamefont {Manzoni}}, \bibinfo {author} {\bibfnamefont
  {D.}~\bibnamefont {Brida}}, \bibinfo {author} {\bibfnamefont
  {G.}~\bibnamefont {Cerullo}}, \ and\ \bibinfo {author} {\bibfnamefont
  {D.}~\bibnamefont {Mihailovic}},\ }\href {\doibase
  10.1103/PhysRevLett.105.257001} {\bibfield  {journal} {\bibinfo  {journal}
  {Phys. Rev. Lett.}\ }\textbf {\bibinfo {volume} {105}},\ \bibinfo {pages}
  {257001} (\bibinfo {year} {2010})}\BibitemShut {NoStop}%
\bibitem [{\citenamefont {Giannetti}\ \emph {et~al.}(2011)\citenamefont
  {Giannetti}, \citenamefont {Cilento}, \citenamefont {Dal~Conte},
  \citenamefont {Coslovich}, \citenamefont {Ferrini}, \citenamefont
  {Molegraaf}, \citenamefont {Raichle}, \citenamefont {Liang}, \citenamefont
  {Eisaki}, \citenamefont {Greven} \emph {et~al.}}]{GiannettiCilento2011}%
  \BibitemOpen
  \bibfield  {author} {\bibinfo {author} {\bibfnamefont {C.}~\bibnamefont
  {Giannetti}}, \bibinfo {author} {\bibfnamefont {F.}~\bibnamefont {Cilento}},
  \bibinfo {author} {\bibfnamefont {S.}~\bibnamefont {Dal~Conte}}, \bibinfo
  {author} {\bibfnamefont {G.}~\bibnamefont {Coslovich}}, \bibinfo {author}
  {\bibfnamefont {G.}~\bibnamefont {Ferrini}}, \bibinfo {author} {\bibfnamefont
  {H.}~\bibnamefont {Molegraaf}}, \bibinfo {author} {\bibfnamefont
  {M.}~\bibnamefont {Raichle}}, \bibinfo {author} {\bibfnamefont
  {R.}~\bibnamefont {Liang}}, \bibinfo {author} {\bibfnamefont
  {H.}~\bibnamefont {Eisaki}}, \bibinfo {author} {\bibfnamefont
  {M.}~\bibnamefont {Greven}},  \emph {et~al.},\ }\href@noop {} {\bibfield
  {journal} {\bibinfo  {journal} {Nature Communications}\ }\textbf {\bibinfo
  {volume} {2}},\ \bibinfo {pages} {353} (\bibinfo {year} {2011})}\BibitemShut
  {NoStop}%
\bibitem [{\citenamefont {Coslovich}\ \emph {et~al.}(2013)\citenamefont
  {Coslovich}, \citenamefont {Giannetti}, \citenamefont {Cilento},
  \citenamefont {Dal~Conte}, \citenamefont {Abebaw}, \citenamefont {Bossini},
  \citenamefont {Ferrini}, \citenamefont {Eisaki}, \citenamefont {Greven},
  \citenamefont {Damascelli},\ and\ \citenamefont
  {Parmigiani}}]{CoslovichGiannetti2013}%
  \BibitemOpen
  \bibfield  {author} {\bibinfo {author} {\bibfnamefont {G.}~\bibnamefont
  {Coslovich}}, \bibinfo {author} {\bibfnamefont {C.}~\bibnamefont
  {Giannetti}}, \bibinfo {author} {\bibfnamefont {F.}~\bibnamefont {Cilento}},
  \bibinfo {author} {\bibfnamefont {S.}~\bibnamefont {Dal~Conte}}, \bibinfo
  {author} {\bibfnamefont {T.}~\bibnamefont {Abebaw}}, \bibinfo {author}
  {\bibfnamefont {D.}~\bibnamefont {Bossini}}, \bibinfo {author} {\bibfnamefont
  {G.}~\bibnamefont {Ferrini}}, \bibinfo {author} {\bibfnamefont
  {H.}~\bibnamefont {Eisaki}}, \bibinfo {author} {\bibfnamefont
  {M.}~\bibnamefont {Greven}}, \bibinfo {author} {\bibfnamefont
  {A.}~\bibnamefont {Damascelli}}, \ and\ \bibinfo {author} {\bibfnamefont
  {F.}~\bibnamefont {Parmigiani}},\ }\href {\doibase
  10.1103/PhysRevLett.110.107003} {\bibfield  {journal} {\bibinfo  {journal}
  {Phys. Rev. Lett.}\ }\textbf {\bibinfo {volume} {110}},\ \bibinfo {pages}
  {107003} (\bibinfo {year} {2013})}\BibitemShut {NoStop}%
\bibitem [{\citenamefont {Kim}\ \emph {et~al.}(2012)\citenamefont {Kim},
  \citenamefont {Pashkin}, \citenamefont {Sch{\"a}fer}, \citenamefont {Beyer},
  \citenamefont {Porer}, \citenamefont {Wolf}, \citenamefont {Bernhard},
  \citenamefont {Demsar}, \citenamefont {Huber},\ and\ \citenamefont
  {Leitenstorfer}}]{KimPashkin2012}%
  \BibitemOpen
  \bibfield  {author} {\bibinfo {author} {\bibfnamefont {K.~W.}\ \bibnamefont
  {Kim}}, \bibinfo {author} {\bibfnamefont {A.}~\bibnamefont {Pashkin}},
  \bibinfo {author} {\bibfnamefont {H.}~\bibnamefont {Sch{\"a}fer}}, \bibinfo
  {author} {\bibfnamefont {M.}~\bibnamefont {Beyer}}, \bibinfo {author}
  {\bibfnamefont {M.}~\bibnamefont {Porer}}, \bibinfo {author} {\bibfnamefont
  {T.}~\bibnamefont {Wolf}}, \bibinfo {author} {\bibfnamefont {C.}~\bibnamefont
  {Bernhard}}, \bibinfo {author} {\bibfnamefont {J.}~\bibnamefont {Demsar}},
  \bibinfo {author} {\bibfnamefont {R.}~\bibnamefont {Huber}}, \ and\ \bibinfo
  {author} {\bibfnamefont {A.}~\bibnamefont {Leitenstorfer}},\ }\href@noop {}
  {\bibfield  {journal} {\bibinfo  {journal} {Nature Materials}\ }\textbf
  {\bibinfo {volume} {11}},\ \bibinfo {pages} {497} (\bibinfo {year}
  {2012})}\BibitemShut {NoStop}%
\bibitem [{\citenamefont {Stojchevska}\ \emph {et~al.}(2010)\citenamefont
  {Stojchevska}, \citenamefont {Kusar}, \citenamefont {Mertelj}, \citenamefont
  {Kabanov}, \citenamefont {Lin}, \citenamefont {Cao}, \citenamefont {Xu},\
  and\ \citenamefont {Mihailovic}}]{StojchevskaKusar2010}%
  \BibitemOpen
  \bibfield  {author} {\bibinfo {author} {\bibfnamefont {L.}~\bibnamefont
  {Stojchevska}}, \bibinfo {author} {\bibfnamefont {P.}~\bibnamefont {Kusar}},
  \bibinfo {author} {\bibfnamefont {T.}~\bibnamefont {Mertelj}}, \bibinfo
  {author} {\bibfnamefont {V.}~\bibnamefont {Kabanov}}, \bibinfo {author}
  {\bibfnamefont {X.}~\bibnamefont {Lin}}, \bibinfo {author} {\bibfnamefont
  {G.}~\bibnamefont {Cao}}, \bibinfo {author} {\bibfnamefont {Z.}~\bibnamefont
  {Xu}}, \ and\ \bibinfo {author} {\bibfnamefont {D.}~\bibnamefont
  {Mihailovic}},\ }\href@noop {} {\bibfield  {journal} {\bibinfo  {journal}
  {Arxiv preprint arXiv:1002.2582}\ } (\bibinfo {year} {2010})}\BibitemShut
  {NoStop}%
\bibitem [{\citenamefont {Rettig}\ \emph {et~al.}(2012)\citenamefont {Rettig},
  \citenamefont {Cort\'es}, \citenamefont {Thirupathaiah}, \citenamefont
  {Gegenwart}, \citenamefont {Jeevan}, \citenamefont {Wolf}, \citenamefont
  {Fink},\ and\ \citenamefont {Bovensiepen}}]{RettigCortes2012}%
  \BibitemOpen
  \bibfield  {author} {\bibinfo {author} {\bibfnamefont {L.}~\bibnamefont
  {Rettig}}, \bibinfo {author} {\bibfnamefont {R.}~\bibnamefont {Cort\'es}},
  \bibinfo {author} {\bibfnamefont {S.}~\bibnamefont {Thirupathaiah}}, \bibinfo
  {author} {\bibfnamefont {P.}~\bibnamefont {Gegenwart}}, \bibinfo {author}
  {\bibfnamefont {H.~S.}\ \bibnamefont {Jeevan}}, \bibinfo {author}
  {\bibfnamefont {M.}~\bibnamefont {Wolf}}, \bibinfo {author} {\bibfnamefont
  {J.}~\bibnamefont {Fink}}, \ and\ \bibinfo {author} {\bibfnamefont
  {U.}~\bibnamefont {Bovensiepen}},\ }\href {\doibase
  10.1103/PhysRevLett.108.097002} {\bibfield  {journal} {\bibinfo  {journal}
  {Phys. Rev. Lett.}\ }\textbf {\bibinfo {volume} {108}},\ \bibinfo {pages}
  {097002} (\bibinfo {year} {2012})}\BibitemShut {NoStop}%
\bibitem [{\citenamefont {Mertelj}\ \emph {et~al.}(2010)\citenamefont
  {Mertelj}, \citenamefont {Kusar}, \citenamefont {Kabanov}, \citenamefont
  {Stojchevska}, \citenamefont {Zhigadlo}, \citenamefont {Katrych},
  \citenamefont {Bukowski}, \citenamefont {Karpinski}, \citenamefont
  {Weyeneth},\ and\ \citenamefont {Mihailovic}}]{MerteljKusar2010}%
  \BibitemOpen
  \bibfield  {author} {\bibinfo {author} {\bibfnamefont {T.}~\bibnamefont
  {Mertelj}}, \bibinfo {author} {\bibfnamefont {P.}~\bibnamefont {Kusar}},
  \bibinfo {author} {\bibfnamefont {V.~V.}\ \bibnamefont {Kabanov}}, \bibinfo
  {author} {\bibfnamefont {L.}~\bibnamefont {Stojchevska}}, \bibinfo {author}
  {\bibfnamefont {N.~D.}\ \bibnamefont {Zhigadlo}}, \bibinfo {author}
  {\bibfnamefont {S.}~\bibnamefont {Katrych}}, \bibinfo {author} {\bibfnamefont
  {Z.}~\bibnamefont {Bukowski}}, \bibinfo {author} {\bibfnamefont
  {J.}~\bibnamefont {Karpinski}}, \bibinfo {author} {\bibfnamefont
  {S.}~\bibnamefont {Weyeneth}}, \ and\ \bibinfo {author} {\bibfnamefont
  {D.}~\bibnamefont {Mihailovic}},\ }\href {\doibase
  10.1103/PhysRevB.81.224504} {\bibfield  {journal} {\bibinfo  {journal} {Phys.
  Rev. B}\ }\textbf {\bibinfo {volume} {81}},\ \bibinfo {pages} {224504}
  (\bibinfo {year} {2010})}\BibitemShut {NoStop}%
\bibitem [{\citenamefont {Sch\"afer}\ \emph {et~al.}(2010)\citenamefont
  {Sch\"afer}, \citenamefont {Kabanov}, \citenamefont {Beyer}, \citenamefont
  {Biljakovic},\ and\ \citenamefont {Demsar}}]{SchaeferKabanov2010}%
  \BibitemOpen
  \bibfield  {author} {\bibinfo {author} {\bibfnamefont {H.}~\bibnamefont
  {Sch\"afer}}, \bibinfo {author} {\bibfnamefont {V.~V.}\ \bibnamefont
  {Kabanov}}, \bibinfo {author} {\bibfnamefont {M.}~\bibnamefont {Beyer}},
  \bibinfo {author} {\bibfnamefont {K.}~\bibnamefont {Biljakovic}}, \ and\
  \bibinfo {author} {\bibfnamefont {J.}~\bibnamefont {Demsar}},\ }\href
  {\doibase 10.1103/PhysRevLett.105.066402} {\bibfield  {journal} {\bibinfo
  {journal} {Phys. Rev. Lett.}\ }\textbf {\bibinfo {volume} {105}},\ \bibinfo
  {pages} {066402} (\bibinfo {year} {2010})}\BibitemShut {NoStop}%
\bibitem [{\citenamefont {Jiao}\ \emph {et~al.}(2011)\citenamefont {Jiao},
  \citenamefont {Tao}, \citenamefont {Bao}, \citenamefont {Sun}, \citenamefont
  {Feng}, \citenamefont {Xu}, \citenamefont {Nowik}, \citenamefont {Felner},\
  and\ \citenamefont {Cao}}]{JiaoTao2011}%
  \BibitemOpen
  \bibfield  {author} {\bibinfo {author} {\bibfnamefont {W.-H.}\ \bibnamefont
  {Jiao}}, \bibinfo {author} {\bibfnamefont {Q.}~\bibnamefont {Tao}}, \bibinfo
  {author} {\bibfnamefont {J.-K.}\ \bibnamefont {Bao}}, \bibinfo {author}
  {\bibfnamefont {Y.-L.}\ \bibnamefont {Sun}}, \bibinfo {author} {\bibfnamefont
  {C.-M.}\ \bibnamefont {Feng}}, \bibinfo {author} {\bibfnamefont {Z.-A.}\
  \bibnamefont {Xu}}, \bibinfo {author} {\bibfnamefont {I.}~\bibnamefont
  {Nowik}}, \bibinfo {author} {\bibfnamefont {I.}~\bibnamefont {Felner}}, \
  and\ \bibinfo {author} {\bibfnamefont {G.-H.}\ \bibnamefont {Cao}},\
  }\href@noop {} {\bibfield  {journal} {\bibinfo  {journal} {EPL (Europhysics
  Letters)}\ }\textbf {\bibinfo {volume} {95}},\ \bibinfo {pages} {67007}
  (\bibinfo {year} {2011})}\BibitemShut {NoStop}%
\bibitem [{\citenamefont {Chu}\ \emph {et~al.}(2009)\citenamefont {Chu},
  \citenamefont {Analytis}, \citenamefont {Kucharczyk},\ and\ \citenamefont
  {Fisher}}]{ChuAnalytis2009}%
  \BibitemOpen
  \bibfield  {author} {\bibinfo {author} {\bibfnamefont {J.-H.}\ \bibnamefont
  {Chu}}, \bibinfo {author} {\bibfnamefont {J.~G.}\ \bibnamefont {Analytis}},
  \bibinfo {author} {\bibfnamefont {C.}~\bibnamefont {Kucharczyk}}, \ and\
  \bibinfo {author} {\bibfnamefont {I.~R.}\ \bibnamefont {Fisher}},\ }\href
  {\doibase 10.1103/PhysRevB.79.014506} {\bibfield  {journal} {\bibinfo
  {journal} {Phys. Rev. B}\ }\textbf {\bibinfo {volume} {79}},\ \bibinfo
  {pages} {014506} (\bibinfo {year} {2009})}\BibitemShut {NoStop}%
\bibitem [{\citenamefont {Tegel}\ \emph {et~al.}(2008)\citenamefont {Tegel},
  \citenamefont {Rotter}, \citenamefont {Wei{\ss}}, \citenamefont
  {Schappacher}, \citenamefont {P{\"o}ttgen},\ and\ \citenamefont
  {Johrendt}}]{tegelRotter2008}%
  \BibitemOpen
  \bibfield  {author} {\bibinfo {author} {\bibfnamefont {M.}~\bibnamefont
  {Tegel}}, \bibinfo {author} {\bibfnamefont {M.}~\bibnamefont {Rotter}},
  \bibinfo {author} {\bibfnamefont {V.}~\bibnamefont {Wei{\ss}}}, \bibinfo
  {author} {\bibfnamefont {F.~M.}\ \bibnamefont {Schappacher}}, \bibinfo
  {author} {\bibfnamefont {R.}~\bibnamefont {P{\"o}ttgen}}, \ and\ \bibinfo
  {author} {\bibfnamefont {D.}~\bibnamefont {Johrendt}},\ }\href@noop {}
  {\bibfield  {journal} {\bibinfo  {journal} {Journal of Physics: Condensed
  Matter}\ }\textbf {\bibinfo {volume} {20}},\ \bibinfo {pages} {452201}
  (\bibinfo {year} {2008})}\BibitemShut {NoStop}%
\bibitem [{\citenamefont {Charnukha}\ \emph {et~al.}(2013)\citenamefont
  {Charnukha}, \citenamefont {Pr\"opper}, \citenamefont {Larkin}, \citenamefont
  {Sun}, \citenamefont {Li}, \citenamefont {Lin}, \citenamefont {Wolf},
  \citenamefont {Keimer},\ and\ \citenamefont {Boris}}]{CharnukhaLarkin2013}%
  \BibitemOpen
  \bibfield  {author} {\bibinfo {author} {\bibfnamefont {A.}~\bibnamefont
  {Charnukha}}, \bibinfo {author} {\bibfnamefont {D.}~\bibnamefont
  {Pr\"opper}}, \bibinfo {author} {\bibfnamefont {T.~I.}\ \bibnamefont
  {Larkin}}, \bibinfo {author} {\bibfnamefont {D.~L.}\ \bibnamefont {Sun}},
  \bibinfo {author} {\bibfnamefont {Z.~W.}\ \bibnamefont {Li}}, \bibinfo
  {author} {\bibfnamefont {C.~T.}\ \bibnamefont {Lin}}, \bibinfo {author}
  {\bibfnamefont {T.}~\bibnamefont {Wolf}}, \bibinfo {author} {\bibfnamefont
  {B.}~\bibnamefont {Keimer}}, \ and\ \bibinfo {author} {\bibfnamefont {A.~V.}\
  \bibnamefont {Boris}},\ }\href {\doibase 10.1103/PhysRevB.88.184511}
  {\bibfield  {journal} {\bibinfo  {journal} {Phys. Rev. B}\ }\textbf {\bibinfo
  {volume} {88}},\ \bibinfo {pages} {184511} (\bibinfo {year}
  {2013})}\BibitemShut {NoStop}%
\bibitem [{Note1()}]{Note1}%
  \BibitemOpen
  \bibinfo {note} {For Sr-122 and Eu-122 we performed SVD decomposition joining
  together the data for both polarizations at each $T$.}\BibitemShut {Stop}%
\bibitem [{Note2()}]{Note2}%
  \BibitemOpen
  \bibinfo {note} {Somewhat larger $\sigma $ in Ba-122 and Eu-122 might be
  attributed to a slightly worse chirp compensation.}\BibitemShut {Stop}%
\bibitem [{\citenamefont {Dvorsek}\ \emph {et~al.}(2002)\citenamefont
  {Dvorsek}, \citenamefont {Kabanov}, \citenamefont {Demsar}, \citenamefont
  {Kazakov}, \citenamefont {Karpinski},\ and\ \citenamefont
  {Mihailovic}}]{DvorsekKabanov2002}%
  \BibitemOpen
  \bibfield  {author} {\bibinfo {author} {\bibfnamefont {D.}~\bibnamefont
  {Dvorsek}}, \bibinfo {author} {\bibfnamefont {V.~V.}\ \bibnamefont
  {Kabanov}}, \bibinfo {author} {\bibfnamefont {J.}~\bibnamefont {Demsar}},
  \bibinfo {author} {\bibfnamefont {S.~M.}\ \bibnamefont {Kazakov}}, \bibinfo
  {author} {\bibfnamefont {J.}~\bibnamefont {Karpinski}}, \ and\ \bibinfo
  {author} {\bibfnamefont {D.}~\bibnamefont {Mihailovic}},\ }\href {\doibase
  10.1103/PhysRevB.66.020510} {\bibfield  {journal} {\bibinfo  {journal} {Phys.
  Rev. B}\ }\textbf {\bibinfo {volume} {66}},\ \bibinfo {pages} {020510}
  (\bibinfo {year} {2002})}\BibitemShut {NoStop}%
\bibitem [{\citenamefont {Yi}\ \emph {et~al.}(2011)\citenamefont {Yi},
  \citenamefont {Lu}, \citenamefont {Chu}, \citenamefont {Analytis},
  \citenamefont {Sorini}, \citenamefont {Kemper}, \citenamefont {Moritz},
  \citenamefont {Mo}, \citenamefont {Moore}, \citenamefont {Hashimoto},
  \citenamefont {Lee}, \citenamefont {Hussain}, \citenamefont {Devereaux},
  \citenamefont {Fisher},\ and\ \citenamefont {Shen}}]{YiLu2011}%
  \BibitemOpen
  \bibfield  {author} {\bibinfo {author} {\bibfnamefont {M.}~\bibnamefont
  {Yi}}, \bibinfo {author} {\bibfnamefont {D.}~\bibnamefont {Lu}}, \bibinfo
  {author} {\bibfnamefont {J.-H.}\ \bibnamefont {Chu}}, \bibinfo {author}
  {\bibfnamefont {J.~G.}\ \bibnamefont {Analytis}}, \bibinfo {author}
  {\bibfnamefont {A.~P.}\ \bibnamefont {Sorini}}, \bibinfo {author}
  {\bibfnamefont {A.~F.}\ \bibnamefont {Kemper}}, \bibinfo {author}
  {\bibfnamefont {B.}~\bibnamefont {Moritz}}, \bibinfo {author} {\bibfnamefont
  {S.-K.}\ \bibnamefont {Mo}}, \bibinfo {author} {\bibfnamefont {R.~G.}\
  \bibnamefont {Moore}}, \bibinfo {author} {\bibfnamefont {M.}~\bibnamefont
  {Hashimoto}}, \bibinfo {author} {\bibfnamefont {W.-S.}\ \bibnamefont {Lee}},
  \bibinfo {author} {\bibfnamefont {Z.}~\bibnamefont {Hussain}}, \bibinfo
  {author} {\bibfnamefont {T.~P.}\ \bibnamefont {Devereaux}}, \bibinfo {author}
  {\bibfnamefont {I.~R.}\ \bibnamefont {Fisher}}, \ and\ \bibinfo {author}
  {\bibfnamefont {Z.-X.}\ \bibnamefont {Shen}},\ }\href {\doibase
  10.1073/pnas.1015572108} {\bibfield  {journal} {\bibinfo  {journal}
  {Proceedings of the National Academy of Sciences}\ }\textbf {\bibinfo
  {volume} {108}},\ \bibinfo {pages} {6878} (\bibinfo {year}
  {2011})}\BibitemShut {NoStop}%
\bibitem [{\citenamefont {Analytis}\ \emph {et~al.}(2009)\citenamefont
  {Analytis}, \citenamefont {McDonald}, \citenamefont {Chu}, \citenamefont
  {Riggs}, \citenamefont {Bangura}, \citenamefont {Kucharczyk}, \citenamefont
  {Johannes},\ and\ \citenamefont {Fisher}}]{AnalytisMcDonald2009}%
  \BibitemOpen
  \bibfield  {author} {\bibinfo {author} {\bibfnamefont {J.}~\bibnamefont
  {Analytis}}, \bibinfo {author} {\bibfnamefont {R.}~\bibnamefont {McDonald}},
  \bibinfo {author} {\bibfnamefont {J.}~\bibnamefont {Chu}}, \bibinfo {author}
  {\bibfnamefont {S.}~\bibnamefont {Riggs}}, \bibinfo {author} {\bibfnamefont
  {A.}~\bibnamefont {Bangura}}, \bibinfo {author} {\bibfnamefont
  {C.}~\bibnamefont {Kucharczyk}}, \bibinfo {author} {\bibfnamefont
  {M.}~\bibnamefont {Johannes}}, \ and\ \bibinfo {author} {\bibfnamefont
  {I.}~\bibnamefont {Fisher}},\ }\href@noop {} {\bibfield  {journal} {\bibinfo
  {journal} {Physical Review B}\ }\textbf {\bibinfo {volume} {80}},\ \bibinfo
  {pages} {64507} (\bibinfo {year} {2009})}\BibitemShut {NoStop}%
\bibitem [{\citenamefont {Hu}\ \emph {et~al.}(2008)\citenamefont {Hu},
  \citenamefont {Dong}, \citenamefont {Li}, \citenamefont {Li}, \citenamefont
  {Zheng}, \citenamefont {Chen}, \citenamefont {Luo},\ and\ \citenamefont
  {Wang}}]{HuDong2008}%
  \BibitemOpen
  \bibfield  {author} {\bibinfo {author} {\bibfnamefont {W.~Z.}\ \bibnamefont
  {Hu}}, \bibinfo {author} {\bibfnamefont {J.}~\bibnamefont {Dong}}, \bibinfo
  {author} {\bibfnamefont {G.}~\bibnamefont {Li}}, \bibinfo {author}
  {\bibfnamefont {Z.}~\bibnamefont {Li}}, \bibinfo {author} {\bibfnamefont
  {P.}~\bibnamefont {Zheng}}, \bibinfo {author} {\bibfnamefont {G.~F.}\
  \bibnamefont {Chen}}, \bibinfo {author} {\bibfnamefont {J.~L.}\ \bibnamefont
  {Luo}}, \ and\ \bibinfo {author} {\bibfnamefont {N.~L.}\ \bibnamefont
  {Wang}},\ }\href {\doibase 10.1103/PhysRevLett.101.257005} {\bibfield
  {journal} {\bibinfo  {journal} {Physical Review Letters}\ }\textbf {\bibinfo
  {volume} {101}},\ \bibinfo {eid} {257005} (\bibinfo {year}
  {2008})}\BibitemShut {NoStop}%
\bibitem [{\citenamefont {Wu}\ \emph {et~al.}(2009)\citenamefont {Wu},
  \citenamefont {Bari\ifmmode \check{s}\else \v{s}\fi{}i\ifmmode~\acute{c}\else
  \'{c}\fi{}}, \citenamefont {Drichko}, \citenamefont {Kaiser}, \citenamefont
  {Faridian}, \citenamefont {Dressel}, \citenamefont {Jiang}, \citenamefont
  {Ren}, \citenamefont {Li}, \citenamefont {Cao}, \citenamefont {Xu},
  \citenamefont {Jeevan},\ and\ \citenamefont {Gegenwart}}]{WuBarisic2009}%
  \BibitemOpen
  \bibfield  {author} {\bibinfo {author} {\bibfnamefont {D.}~\bibnamefont
  {Wu}}, \bibinfo {author} {\bibfnamefont {N.}~\bibnamefont {Bari\ifmmode
  \check{s}\else \v{s}\fi{}i\ifmmode~\acute{c}\else \'{c}\fi{}}}, \bibinfo
  {author} {\bibfnamefont {N.}~\bibnamefont {Drichko}}, \bibinfo {author}
  {\bibfnamefont {S.}~\bibnamefont {Kaiser}}, \bibinfo {author} {\bibfnamefont
  {A.}~\bibnamefont {Faridian}}, \bibinfo {author} {\bibfnamefont
  {M.}~\bibnamefont {Dressel}}, \bibinfo {author} {\bibfnamefont
  {S.}~\bibnamefont {Jiang}}, \bibinfo {author} {\bibfnamefont
  {Z.}~\bibnamefont {Ren}}, \bibinfo {author} {\bibfnamefont {L.~J.}\
  \bibnamefont {Li}}, \bibinfo {author} {\bibfnamefont {G.~H.}\ \bibnamefont
  {Cao}}, \bibinfo {author} {\bibfnamefont {Z.~A.}\ \bibnamefont {Xu}},
  \bibinfo {author} {\bibfnamefont {H.~S.}\ \bibnamefont {Jeevan}}, \ and\
  \bibinfo {author} {\bibfnamefont {P.}~\bibnamefont {Gegenwart}},\ }\href
  {\doibase 10.1103/PhysRevB.79.155103} {\bibfield  {journal} {\bibinfo
  {journal} {Phys. Rev. B}\ }\textbf {\bibinfo {volume} {79}},\ \bibinfo
  {pages} {155103} (\bibinfo {year} {2009})}\BibitemShut {NoStop}%
\bibitem [{\citenamefont {Gadermaier}\ \emph {et~al.}(2012)\citenamefont
  {Gadermaier}, \citenamefont {Kabanov}, \citenamefont {Alexandrov},
  \citenamefont {Stojchevska}, \citenamefont {Mertelj}, \citenamefont
  {Manzoni}, \citenamefont {Cerullo}, \citenamefont {Zhigadlo}, \citenamefont
  {Karpinski}, \citenamefont {Cai} \emph {et~al.}}]{gadermaierKabanov2012}%
  \BibitemOpen
  \bibfield  {author} {\bibinfo {author} {\bibfnamefont {C.}~\bibnamefont
  {Gadermaier}}, \bibinfo {author} {\bibfnamefont {V.}~\bibnamefont {Kabanov}},
  \bibinfo {author} {\bibfnamefont {A.}~\bibnamefont {Alexandrov}}, \bibinfo
  {author} {\bibfnamefont {L.}~\bibnamefont {Stojchevska}}, \bibinfo {author}
  {\bibfnamefont {T.}~\bibnamefont {Mertelj}}, \bibinfo {author} {\bibfnamefont
  {C.}~\bibnamefont {Manzoni}}, \bibinfo {author} {\bibfnamefont
  {G.}~\bibnamefont {Cerullo}}, \bibinfo {author} {\bibfnamefont
  {N.}~\bibnamefont {Zhigadlo}}, \bibinfo {author} {\bibfnamefont
  {J.}~\bibnamefont {Karpinski}}, \bibinfo {author} {\bibfnamefont
  {Y.}~\bibnamefont {Cai}},  \emph {et~al.},\ }\href@noop {} {\bibfield
  {journal} {\bibinfo  {journal} {Arxiv preprint arXiv:1205.4978}\ } (\bibinfo
  {year} {2012})}\BibitemShut {NoStop}%
\bibitem [{\citenamefont {Zbiri}\ \emph {et~al.}(2010)\citenamefont {Zbiri},
  \citenamefont {Mittal}, \citenamefont {Rols}, \citenamefont {Su},
  \citenamefont {Xiao}, \citenamefont {Schober}, \citenamefont {Chaplot},
  \citenamefont {Johnson}, \citenamefont {Chatterji}, \citenamefont {Inoue},
  \citenamefont {Matsuishi}, \citenamefont {Hosono},\ and\ \citenamefont
  {Brueckel}}]{ZbiriMittal2010}%
  \BibitemOpen
  \bibfield  {author} {\bibinfo {author} {\bibfnamefont {M.}~\bibnamefont
  {Zbiri}}, \bibinfo {author} {\bibfnamefont {R.}~\bibnamefont {Mittal}},
  \bibinfo {author} {\bibfnamefont {S.}~\bibnamefont {Rols}}, \bibinfo {author}
  {\bibfnamefont {Y.}~\bibnamefont {Su}}, \bibinfo {author} {\bibfnamefont
  {Y.}~\bibnamefont {Xiao}}, \bibinfo {author} {\bibfnamefont {H.}~\bibnamefont
  {Schober}}, \bibinfo {author} {\bibfnamefont {S.~L.}\ \bibnamefont
  {Chaplot}}, \bibinfo {author} {\bibfnamefont {M.~R.}\ \bibnamefont
  {Johnson}}, \bibinfo {author} {\bibfnamefont {T.}~\bibnamefont {Chatterji}},
  \bibinfo {author} {\bibfnamefont {Y.}~\bibnamefont {Inoue}}, \bibinfo
  {author} {\bibfnamefont {S.}~\bibnamefont {Matsuishi}}, \bibinfo {author}
  {\bibfnamefont {H.}~\bibnamefont {Hosono}}, \ and\ \bibinfo {author}
  {\bibfnamefont {T.}~\bibnamefont {Brueckel}},\ }\href
  {http://stacks.iop.org/0953-8984/22/i=31/a=315701} {\bibfield  {journal}
  {\bibinfo  {journal} {Journal of Physics: Condensed Matter}\ }\textbf
  {\bibinfo {volume} {22}},\ \bibinfo {pages} {315701} (\bibinfo {year}
  {2010})}\BibitemShut {NoStop}%
\bibitem [{\citenamefont {Ewings}\ \emph {et~al.}(2008)\citenamefont {Ewings},
  \citenamefont {Perring}, \citenamefont {Bewley}, \citenamefont {Guidi},
  \citenamefont {Pitcher}, \citenamefont {Parker}, \citenamefont {Clarke},\
  and\ \citenamefont {Boothroyd}}]{EwingsPerring2008}%
  \BibitemOpen
  \bibfield  {author} {\bibinfo {author} {\bibfnamefont {R.~A.}\ \bibnamefont
  {Ewings}}, \bibinfo {author} {\bibfnamefont {T.~G.}\ \bibnamefont {Perring}},
  \bibinfo {author} {\bibfnamefont {R.~I.}\ \bibnamefont {Bewley}}, \bibinfo
  {author} {\bibfnamefont {T.}~\bibnamefont {Guidi}}, \bibinfo {author}
  {\bibfnamefont {M.~J.}\ \bibnamefont {Pitcher}}, \bibinfo {author}
  {\bibfnamefont {D.~R.}\ \bibnamefont {Parker}}, \bibinfo {author}
  {\bibfnamefont {S.~J.}\ \bibnamefont {Clarke}}, \ and\ \bibinfo {author}
  {\bibfnamefont {A.~T.}\ \bibnamefont {Boothroyd}},\ }\href {\doibase
  10.1103/PhysRevB.78.220501} {\bibfield  {journal} {\bibinfo  {journal} {Phys.
  Rev. B}\ }\textbf {\bibinfo {volume} {78}},\ \bibinfo {pages} {220501}
  (\bibinfo {year} {2008})}\BibitemShut {NoStop}%
\bibitem [{\citenamefont {Ewings}\ \emph {et~al.}(2011)\citenamefont {Ewings},
  \citenamefont {Perring}, \citenamefont {Gillett}, \citenamefont {Das},
  \citenamefont {Sebastian}, \citenamefont {Taylor}, \citenamefont {Guidi},\
  and\ \citenamefont {Boothroyd}}]{EwingsPerring2011}%
  \BibitemOpen
  \bibfield  {author} {\bibinfo {author} {\bibfnamefont {R.~A.}\ \bibnamefont
  {Ewings}}, \bibinfo {author} {\bibfnamefont {T.~G.}\ \bibnamefont {Perring}},
  \bibinfo {author} {\bibfnamefont {J.}~\bibnamefont {Gillett}}, \bibinfo
  {author} {\bibfnamefont {S.~D.}\ \bibnamefont {Das}}, \bibinfo {author}
  {\bibfnamefont {S.~E.}\ \bibnamefont {Sebastian}}, \bibinfo {author}
  {\bibfnamefont {A.~E.}\ \bibnamefont {Taylor}}, \bibinfo {author}
  {\bibfnamefont {T.}~\bibnamefont {Guidi}}, \ and\ \bibinfo {author}
  {\bibfnamefont {A.~T.}\ \bibnamefont {Boothroyd}},\ }\href {\doibase
  10.1103/PhysRevB.83.214519} {\bibfield  {journal} {\bibinfo  {journal} {Phys.
  Rev. B}\ }\textbf {\bibinfo {volume} {83}},\ \bibinfo {pages} {214519}
  (\bibinfo {year} {2011})}\BibitemShut {NoStop}%
\bibitem [{\citenamefont {Sugai}\ \emph {et~al.}(2013)\citenamefont {Sugai},
  \citenamefont {Mizuno}, \citenamefont {Watanabe}, \citenamefont {Kawaguchi},
  \citenamefont {Takenaka}, \citenamefont {Ikuta}, \citenamefont {Kiho},
  \citenamefont {Nakajima}, \citenamefont {Lee}, \citenamefont {Iyo},
  \citenamefont {Eisaki},\ and\ \citenamefont {Uchida}}]{SugaiMizuno2013}%
  \BibitemOpen
  \bibfield  {author} {\bibinfo {author} {\bibfnamefont {S.}~\bibnamefont
  {Sugai}}, \bibinfo {author} {\bibfnamefont {Y.}~\bibnamefont {Mizuno}},
  \bibinfo {author} {\bibfnamefont {R.}~\bibnamefont {Watanabe}}, \bibinfo
  {author} {\bibfnamefont {T.}~\bibnamefont {Kawaguchi}}, \bibinfo {author}
  {\bibfnamefont {K.}~\bibnamefont {Takenaka}}, \bibinfo {author}
  {\bibfnamefont {H.}~\bibnamefont {Ikuta}}, \bibinfo {author} {\bibfnamefont
  {K.}~\bibnamefont {Kiho}}, \bibinfo {author} {\bibfnamefont {M.}~\bibnamefont
  {Nakajima}}, \bibinfo {author} {\bibfnamefont {C.}~\bibnamefont {Lee}},
  \bibinfo {author} {\bibfnamefont {A.}~\bibnamefont {Iyo}}, \bibinfo {author}
  {\bibfnamefont {H.}~\bibnamefont {Eisaki}}, \ and\ \bibinfo {author}
  {\bibfnamefont {S.}~\bibnamefont {Uchida}},\ }\href {\doibase
  10.1007/s10948-012-1966-6} {\bibfield  {journal} {\bibinfo  {journal}
  {Journal of Superconductivity and Novel Magnetism}\ }\textbf {\bibinfo
  {volume} {26}},\ \bibinfo {pages} {1179} (\bibinfo {year}
  {2013})}\BibitemShut {NoStop}%
\bibitem [{\citenamefont {Schaefer}\ \emph {et~al.}(2014)\citenamefont
  {Schaefer}, \citenamefont {Kabanov},\ and\ \citenamefont
  {Demsar}}]{SchaeferKabanov2014}%
  \BibitemOpen
  \bibfield  {author} {\bibinfo {author} {\bibfnamefont {H.}~\bibnamefont
  {Schaefer}}, \bibinfo {author} {\bibfnamefont {V.~V.}\ \bibnamefont
  {Kabanov}}, \ and\ \bibinfo {author} {\bibfnamefont {J.}~\bibnamefont
  {Demsar}},\ }\href {\doibase 10.1103/PhysRevB.89.045106} {\bibfield
  {journal} {\bibinfo  {journal} {Phys. Rev. B}\ }\textbf {\bibinfo {volume}
  {89}},\ \bibinfo {pages} {045106} (\bibinfo {year} {2014})}\BibitemShut
  {NoStop}%
\bibitem [{\citenamefont {Psaltakis}(1984)}]{psaltakis1984}%
  \BibitemOpen
  \bibfield  {author} {\bibinfo {author} {\bibfnamefont {G.}~\bibnamefont
  {Psaltakis}},\ }\href@noop {} {\bibfield  {journal} {\bibinfo  {journal}
  {Solid state communications}\ }\textbf {\bibinfo {volume} {51}},\ \bibinfo
  {pages} {535} (\bibinfo {year} {1984})}\BibitemShut {NoStop}%
\bibitem [{\citenamefont {Gr{\"u}ner}(1994)}]{gruner1994}%
  \BibitemOpen
  \bibfield  {author} {\bibinfo {author} {\bibfnamefont {G.}~\bibnamefont
  {Gr{\"u}ner}},\ }\href@noop {} {\bibfield  {journal} {\bibinfo  {journal}
  {Reviews of modern physics}\ }\textbf {\bibinfo {volume} {66}},\ \bibinfo
  {pages} {1} (\bibinfo {year} {1994})}\BibitemShut {NoStop}%
\bibitem [{Note3()}]{Note3}%
  \BibitemOpen
  \bibinfo {note} {Since the transition-induced reflectivity change would be
  hardly explained just by $T$-induced change of the quasiparticle distribution
  function at the experimental\cite {CharnukhaLarkin2013} 15-20 K $T$
  difference, it should correspond to the characteristic
  order-parameter-induced spectral change.}\BibitemShut {Stop}%
\bibitem [{\citenamefont {Mertelj}\ \emph {et~al.}(2013)\citenamefont
  {Mertelj}, \citenamefont {Stojchevska}, \citenamefont {Zhigadlo},
  \citenamefont {Karpinski},\ and\ \citenamefont
  {Mihailovic}}]{MerteljStojchevska2013}%
  \BibitemOpen
  \bibfield  {author} {\bibinfo {author} {\bibfnamefont {T.}~\bibnamefont
  {Mertelj}}, \bibinfo {author} {\bibfnamefont {L.}~\bibnamefont
  {Stojchevska}}, \bibinfo {author} {\bibfnamefont {N.~D.}\ \bibnamefont
  {Zhigadlo}}, \bibinfo {author} {\bibfnamefont {J.}~\bibnamefont {Karpinski}},
  \ and\ \bibinfo {author} {\bibfnamefont {D.}~\bibnamefont {Mihailovic}},\
  }\href {\doibase 10.1103/PhysRevB.87.174525} {\bibfield  {journal} {\bibinfo
  {journal} {Phys. Rev. B}\ }\textbf {\bibinfo {volume} {87}},\ \bibinfo
  {pages} {174525} (\bibinfo {year} {2013})}\BibitemShut {NoStop}%
\bibitem [{\citenamefont {Elesin}\ and\ \citenamefont
  {Kopaev}(1981)}]{elesinKopaev1981}%
  \BibitemOpen
  \bibfield  {author} {\bibinfo {author} {\bibfnamefont {V.~F.}\ \bibnamefont
  {Elesin}}\ and\ \bibinfo {author} {\bibfnamefont {Y.~V.}\ \bibnamefont
  {Kopaev}},\ }\href@noop {} {\bibfield  {journal} {\bibinfo  {journal}
  {Physics-Uspekhi}\ }\textbf {\bibinfo {volume} {24}},\ \bibinfo {pages} {116}
  (\bibinfo {year} {1981})}\BibitemShut {NoStop}%
\end{thebibliography}%

\end{document}